\documentclass[12pt]{article}
\usepackage[utf8]{inputenc}
\pdfoutput=1
\usepackage{amsmath,amsfonts,amsthm,mathrsfs} 
\usepackage{float}
\usepackage{mathtools}
\usepackage[ruled,linesnumbered,noend]{algorithm2e} 
\usepackage{color}
\usepackage{bm}
\usepackage{lineno}
\usepackage{graphicx}
\usepackage{hyperref}
\usepackage{caption}
\usepackage{subcaption} 
\usepackage{cases} 
\usepackage[margin=1.5in]{geometry}
\usepackage{setspace}
\hypersetup{colorlinks,urlcolor=blue,citecolor=black}

\usepackage{pgfplots} 
\pgfplotsset{compat=1.16}
\usepackage{geometry} 

\usepackage{color}
\definecolor{forest-green}{rgb}{0.13,0.54,0.13}

\usepackage{soul}
\usepackage{comment}

\usepackage{tikz-qtree}
\usetikzlibrary{arrows}
\usetikzlibrary{patterns}

\usepackage{multirow}
\usepackage{booktabs}

\usepackage{titlesec}
\titleformat{\paragraph}{\normalfont\normalsize\bfseries}{\theparagraph}{1em}{}
\titlespacing*{\paragraph}{0pt}{3.25ex plus 1ex minus .2ex}{1.5ex plus .2ex}

\DeclarePairedDelimiter\ceil{\lceil}{\rceil}
\DeclarePairedDelimiter\floor{\lfloor}{\rfloor}

\newcommand{\Bern}{\text{Bern}}

\newcommand{\I}{\mathbb{I}}
\newcommand{\Lik}{\mathcal{L}}
\newcommand{\M}{\mathbf{M}}

\newcommand{\T}{\mathcal{T}}

\newcommand{\X}{\mathcal{X}}

\newcommand{\E}{\mathbb{E}}

\newcommand{\R}{\mathbb{R}}
\newcommand{\Z}{\mathbb{Z}}

\newcommand{\bfx}{\mathbf{x}}

\newcommand{\bfM}{\mathbf{M}}
\newcommand{\bfX}{\mathbf{X}}
\newcommand{\bfY}{\mathbf{Y}}
\newcommand{\bfZ}{\mathbf{Z}}

\newcommand{\bftheta}{\boldsymbol{\theta}}

\newcommand{\giv}{\ |\ }

\newcommand{\simiid}{\stackrel{iid}{\sim}}
\newcommand{\simind}{\stackrel{\text{ind.}}{\sim}}

\DeclareMathOperator*{\argmax}{arg\,max}
\DeclareMathOperator*{\argmin}{arg\,min}

\SetCommentSty{mycommfont}
\newcommand{\footremember}[2]{%
    \footnote{#2}
    \newcounter{#1}
    \setcounter{#1}{\value{footnote}}%
}
\newcommand{\footrecall}[1]{%
    \footnotemark[\value{#1}]%
} 

\usepackage[round]{natbib}

\usepackage[toc,page,title]{appendix} 

\title{The Taxicab Sampler: MCMC for Discrete Spaces with Application to Tree Models}
\author{%
  Vincent Geels\footremember{addr1}{Department of Statistics, The Ohio State University}\footremember{e1}{geels.1@osu.edu}%
  \and Matthew T. Pratola\footrecall{addr1} \footremember{e2}{mpratola@stat.osu.edu}%
  \and Radu Herbei\footrecall{addr1} \footremember{e3}{herbei.1@osu.edu}%
  }
\date{February 16, 2022}

\begin{document}

\maketitle

\textbf{Abstract. } Motivated by the problem of exploring discrete but very complex state spaces in Bayesian models, we propose a novel Markov Chain Monte Carlo search algorithm: the \textit{taxicab sampler}. We describe the construction of this sampler and discuss how its interpretation and usage differs from that of standard Metropolis-Hastings as well as the related Hamming ball sampler. The proposed sampling algorithm is then shown to demonstrate substantial improvement in computation time without any loss of efficiency relative to a na{\"i}ve Metropolis-Hastings search in a motivating Bayesian regression tree count model, in which we leverage the discrete state space assumption to construct a novel likelihood function that allows for flexibly describing different mean-variance relationships while preserving parameter interpretability compared to existing likelihood functions for count data.

\section{Introduction}
\label{sec:intro}

Bayesian models often use continuous latent variable formulations in problems involving discrete spaces. This approach stems from the work of \cite{albert_1993}, who handled the problem of analyzing binary and polychotomous data within a Bayesian framework via the introduction of Gaussian latent variables; the resulting closed-form conditional posterior distributions facilitated fast parameter updating. More recent examples of analogous continuous latent variable approaches in other discrete settings (namely modeling count response data) may be found in \cite{ghosh_2006}, who describe a latent approach for parameter updating involving power series likelihoods, and \cite{neelon_2019}, who utilizes P\'{o}lya-Gamma mixtures to obtain closed-form updates in the zero-inflated (ZI) negative binomial regression setting. While powerful, viable implementations of a continuous latent variable mechanism are not always straightforward in certain types of discrete problems. Additionally, these formulations often suffer from the fact that the latent variable dimension grows with the sample size $n$.

An alternative approach is to work with discrete spaces directly; a popular example of this may be seen in Bayesian regression tree models (BRTs) where, for instance, the split rules for continuous predictors are modeled using a discrete candidate set. Still, this choice comes with its own difficulties as the posterior tree space in such models is discrete, complex, and large. The resulting issue of poor mixing in BRTs is a known one in the literature \citep{cgm_1998, denison_1998, cgm_2010}, although recent progress has been made \citep{pratola_2016, mohammadi_2020}.

With these opposing perspectives in mind, we consider the general approach of approximating continuous spaces via discretization, a strategy that allows for the replacement of integrals with sums when marginalization is required and often simplifies the problem of identifying modes in probability distributions over compact sets. \cite{ohagan_1990} leverage this discretization strategy, demonstrating a simple yet practical approach to fitting a Bayesian logistic GLM using improper uniform priors. The conditional posterior distributions of interest are subsequently discretized over a finite grid in order to obtain a closed-form posterior that allows for approximate but straightforward sampling. A more recent example may be found in the machine learning literature for continuous control problems, where \cite{metz_2019} propose discretizing high-dimensional continuous action spaces and sequentially obtain actions that conditionally maximize a target action-value function one dimension at a time.

In this paper, we argue for {\em working with discrete spaces directly} and propose an MCMC algorithm, hereafter referred to as the {\em taxicab sampler} (TC sampler), a nod to the way in which the sampler traverses the state space. The TC sampler extends the Hamming ball sampler (HBS) previously formulated by \cite{titsias_2017} and can be applied in scenarios where the latter algorithm cannot. Our proposed algorithm is readily applicable to high-dimensional discrete spaces where a natural distance exists, e.g. the $L_{\infty}$ distance in our example count model application. Importantly, as with the Hamming ball sampler, the taxicab sampler allows for marginalization over “slices” of the model space, in effect resulting in Gibbs sampling from conditional posterior distributions of interest. A key advantage of the sampler formulation lies in its ability to leverage a computationally advantageous local search approach over a discrete space compared to a more costly global search strategy or an inefficient Metropolis-Hastings (MH) proposal scheme. As shall be seen, the TC sampler delivers improved computational efficiency over a na{\"i}ve MH implementation without compromising inference.

Note that while other ``locally-informed" MCMC samplers have recently been proposed for searching over discrete state spaces \citep{zanella_2020,zhou_2021,grathwohl_2021}, their intended usage is similar to that of the Hamming Ball sampler --- e.g. variable selection, weighted permutations, and/or Ising models. As such they differ substantially from our focus on count models, which typically contain an intrinsic ordering with respect to model dimension not found in these other discrete space problems.

The remaining sections of this paper are organized as follows. In Section \ref{sec:taxicab-sampler} we detail the TC sampler mechanism, its role in a model update algorithm, and discuss considerations for dimension-changing proposals when utilizing the sampler. In Section \ref{sec:the-single-tree-model} we describe a single-tree model designed to handle count data involving a novel count likelihood function utilizing discrete parameter spaces. In Section \ref{sec:runtimeex} we demonstrate the TC sampler’s statistical and computational efficiency in simulation experiments when fitting this single-tree model compared to a straightforward MH approach. Section \ref{sec:discussion} offers discussion and directions for future work. 

\section{The Taxicab Sampler}
\label{sec:taxicab-sampler}

For a sample of size $n$, we work with observed responses $\bfY = \{Y_i \}_{i=1}^n \in \R^n$, along with observed covariate matrix $\bfX = (x_{iv}) \in \R^{n \times p}$, where each $\mathbf{X}_i = (x_{i1},...,x_{ip})$ is the row vector of covariates measured along with the $i$th response $Y_i$. We reserve the use of the notation $\bfx_v,\ v=1,...,p$, to exclusively describe the $v$th covariate dimension of $\bfX$ when discussing certain aspects of tree structures in Section \ref{sec:the-single-tree-model}. In general, we also use the notation $\mathbf{Y}_{-h} = (Y_1,\dots,Y_{h-1},Y_{h+1},\dots,Y_n)$ for some vector $\mathbf{Y} \in \R^n$ and we use the square bracket notation $[Z]$ to denote the distribution of a generic random variable $Z$. 

As described in section \ref{sec:intro}, our focus is on developing technology with which we may fit models depending (in part or in whole) on parameter spaces restricted to countable subsets of $\R$. We assume a Bayesian setup where inference is desired on some discrete parameter vector $\bm{\lambda} \in \Z^B$. Choices other than $\Z^B$ are possible. All other model parameters are collected in the vector $\bm{\theta} \in \Theta$. For the purpose of exposition we assume we work with a joint probability distribution $p(\mathbf{Y},\bm{\theta},\bm{\lambda} \giv \mathbf{X})$, where $\mathbf{X}$ is treated as fixed. The unnormalized joint posterior distribution for $(\bm{\theta},\bm{\lambda})$ factorizes as
\begin{equation}
    \pi(\bm{\theta},\bm{\lambda} \giv \mathbf{Y},\mathbf{X}) \propto p(\mathbf{Y} \giv \bm{\theta},\bm{\lambda},\mathbf{X})\pi(\bm{\theta},\bm{\lambda}).
\end{equation}
To explore the discrete parameter spaces associated with $\bm{\lambda}$ during model-fitting, we introduce a fundamental tool called the taxicab sampler. Our sampler takes motivation from the Hamming ball sampler, \citep{titsias_2017} a generic MCMC sampling procedure for high-dimensional discrete-state models that allows for iterative sampling from ``slices" of the model space that are generated via Hamming balls. These Hamming balls are constructed using the Hamming distance, which is defined between two vectors $\mathbf{w},\mathbf{u}$ as

\begin{equation}
d_H(\mathbf{w},\mathbf{u}) = \sum_b \mathbb{I}\{w_b \neq u_b\}.
\end{equation} 
These balls have finite cardinality in settings where $\mathbf{w}$ and $\mathbf{u}$ take on a finite number of configurations, and in such cases they allow for tractable computation of normalizing constants in order to sample from slices of the target conditional posterior distributions. The result is a powerful yet simple technology that allows for Gibbs step updates on parameters or latent variables in problems that would otherwise require more cumbersome MH updates.

As constructed, the HBS is ideal for performing Bayesian inference via MCMC on binary sequences or matrices \citep{titsias_2017}. However, the Hamming distance is less suitable in other discrete-valued discrete state space settings, since enumerating all set elements generated by a Hamming ball becomes significantly more expensive in non-binary settings and intractable when the probability distribution of one or more random variables in the matrix or sequence places positive probability on infinitely many values.

This point is made plain using a simple univariate example: suppose we have a pmf $p(\bfY \giv \lambda)$ and prior distribution $\pi(\lambda)$ with support on $\mathbb{Z}$, such that $\pi(\lambda \giv \bfY)$ is unavailable in closed form. Introducing a latent variable $U$ to facilitate Hamming ball sampling on $\lambda$ requires construction of an auxiliary distribution $p(U \giv \lambda) = \frac{\mathbb{I}\{U \in \mathcal{H}_m(\lambda) \}}{\left| \mathcal{H}_m(\lambda) \right|}$, where $\left| \cdot \right|$ denotes set cardinality. Now $\mathcal{H}_m(\lambda) = \left\{ U \in \mathbb{Z} : d_H(U,\lambda) \leq m \right\} = \left\{ U \in \mathbb{Z} : U \neq \lambda \right\}$, so that $p(U \giv \lambda)$ amounts to an improper uniform distribution on $\mathbb{Z} \setminus \lambda$ and no longer provides an efficient means to aid in sampling the target posterior distribution $\pi(\lambda \giv \bfY)$.

\subsection{The Taxicab Sampler Algorithm}
\label{ss:ball-sampler-construction}

For discrete state spaces of large or infinite cardinality, the $L_{\infty}$ distance allows for straightforward generation of $L_{\infty}$ neighborhoods of a desired radius, so that marginalization of target distributions over all set elements contained within the neighborhood is tractable. With this idea in mind, we construct in this section a TC sampler that relies on use of sets generated within an $L_{\infty}$ neighborhood, centered at a vector $\mathbf{w}$ with some specified radius $m$, i.e.
\begin{equation}
    \mathcal{N}_m(\mathbf{w}) = \left\{\mathbf{u}: \sup_b |u_b-w_b| \leq m \right\},\ m \geq 0,
    \label{eq:l1neighborhood}
\end{equation}
to construct a transition kernel for an augmented bivariate chain.
Algorithm \ref{algo:gentaxicabsampler} outlines the general TC sampler update algorithm. At its core the TC sampler is a data augmentation scheme requiring the injection of a vector of auxiliary variables $\mathbf{U}$ (possessing the same dimension as $\bm{\lambda}$) into the joint probability model in order to explore slices of the conditional posterior distribution for $\bm{\lambda}$; the sampler construction assumes a distribution for $\mathbf{U}$ that depends only on $\bm{\lambda}$, $p(\mathbf{U} \giv \bm{\lambda})$. This allows us to factorize the augmented joint probability model as
\begin{equation}
    p(\mathbf{Y},\bm{\theta},\bm{\lambda},\mathbf{U} \giv \mathbf{X}) = p(\mathbf{Y},\bm{\theta},\bm{\lambda} \giv \mathbf{X})p(\mathbf{U} \giv \bm{\lambda})\ .
    \label{eq:genaugmodelfact}
\end{equation}
In this manuscript we assume a uniform distribution
\begin{equation}
    p(\mathbf{U} \giv \bm{\lambda}) = \frac{\mathbb{I}\{\mathbf{U} \in \mathcal{N}_{m_{\lambda}}(\bm{\lambda}) \}}{Z_{m_{\lambda}}},
    \label{eq:genauxdistn}
\end{equation}
where $Z_{m_{\lambda}}$ is the cardinality of $\mathcal{N}_{m_{\lambda}}(\bm{\lambda})$; however, this choice of distribution is not critical.
\begin{center}
\begin{algorithm}[H]
 \KwData{Realized observations $(Y_1,\mathbf{X}_1),\dots,(Y_n,\mathbf{X}_n)$, starting values $(\bm{\theta}^{(0)},\bm{\lambda}^{(0)},\mathbf{U}^{(0)})$ }
 \KwResult{Approximate posterior samples drawn from $\pi(\bm{\theta},\bm{\lambda},\mathbf{U} \giv (Y_1,\mathbf{X}_1),\dots,(Y_n,\mathbf{X}_n))$}
 \For{$t \text{ in } 1:N_{mcmc}$ iterations}{
    Update $\bm{\theta}^{(t)} \leftarrow \bm{\theta}^{(t-1)}$ using existing MCMC technology \\
	Draw $\mathbf{U}^{(t)} \sim p(\mathbf{U}^{(t)} \giv \mathcal{N}_{m_{\lambda}}(\bm{\lambda}^{(t-1)})) \equiv p(\mathbf{U}^{(t)} \giv \bm{\lambda}^{(t-1)})$ \\
	Draw $\bm{\lambda}^{(t)} \sim q_{m_{\lambda}}(\bm{\lambda}^{(t)} \giv \mathbf{Y},\bm{\theta}^{(t)},\mathcal{N}_{m_{\lambda}}(\mathbf{U}^{(t)}),\mathbf{X}) \equiv  q_{m_{\lambda}}(\bm{\lambda}^{(t)} \giv \mathbf{Y},\bm{\theta}^{(t)},\mathbf{U}^{(t)},\mathbf{X})$ \\
}
 \caption{The general taxicab sampler update algorithm.}
 \label{algo:gentaxicabsampler}
\end{algorithm}
\end{center}
The key idea of the auxiliary vector $\mathbf{U}$ is that it facilitates a speedier and more efficient exploration of the conditional posterior distribution of $\bm{\lambda}$: given some realization $\mathbf{U} = \mathbf{u}$, we may now compute in closed form the slice of the conditional posterior distribution of $\bm{\lambda}$ captured by $\mathcal{N}_{m_{\lambda}}(\mathbf{u})$ via

\begin{align}
    q_{m_{\lambda}}(\bm{\lambda} \giv \mathbf{Y},\bm{\theta},\mathbf{U}=\mathbf{u},\mathbf{X}) &= \frac{\pi( \bm{\lambda} \giv \mathbf{Y},\bm{\theta},\mathbf{X})\mathbb{I}\{ \bm{\lambda} \in \mathcal{N}_{m_{\lambda}}(\mathbf{u})\} }{ \sum\limits_{\bm{\lambda}^* \in \mathbb{Z}^B}\pi( \bm{\lambda}\mbox{*} \giv  \mathbf{Y},\bm{\theta},\mathbf{X}) \mathbb{I}\{ \bm{\lambda}\mbox{*} \in \mathcal{N}_{m_{\lambda}}(\mathbf{u})\}}  \\
    &= \frac{p(\mathbf{Y},\bm{\theta},\bm{\lambda} \giv \mathbf{X}) \mathbb{I}\{ \bm{\lambda} \in \mathcal{N}_{m_{\lambda}}(\mathbf{u})\}  }{ \sum\limits_{\bm{\lambda}^* \in \mathcal{N}_{m_{\lambda}}(\mathbf{u})}p(\mathbf{Y},\bm{\theta},\bm{\lambda}\mbox{*} \giv \mathbf{X})}.
    \label{eq:taxicabkernel}
\end{align}
It is the tractable normalization constant in \eqref{eq:taxicabkernel} that allows for calculation of these slices of the conditional posterior distribution of $\bm{\lambda}$, as it only requires summation over $Z_{m_\lambda}$ total terms. Thus, in contrast to a MH sampler, there is no ``accept-reject'' mechanism; rather, $\bm{\lambda}$ is \textit{drawn} from $q_{m_\lambda}$ above. By construction, higher-probability regions in these conditional posterior slices are visited more frequently by $q_{m_{\lambda}}$, while the auxiliary proposal distribution \eqref{eq:genauxdistn} encourages random exploration of the augmented space via uniform moves within the generated neighborhood. From this sampler construction we see that the choice of $m_{\lambda}$ controls the trade-off between computational speed and degree of exploration of the target space, with larger choices of radius facilitating better exploration of the target space of interest and smaller choices of radii allowing for faster computation speed.

The complete TC sampler may now be assembled according to the following (ordered) two-part Gibbs step: 
\begin{align}
    &\mbox{1. draw } \mathbf{U} \sim p(\mathbf{U} \giv \bm{\lambda}) \label{eq:taxicabpt1} \\
    &\mbox{2. draw } \bm{\lambda} \sim q_{m_{\lambda}}(\bm{\lambda} \giv \mathbf{Y},\bm{\theta},\mathbf{U},\mathbf{X}). \label{eq:taxicabpt2}
\end{align}

In this two-part procedure, we observe that the TC sampler replaces individual draws of $\bm{\lambda}$ with draws of $(\mathbf{U},\bm{\lambda})$ from the augmented chain
\begin{align*}
  \cdots,
    \begin{pmatrix}
       \mathbf{U}^{(t)} \\
       \bm{\lambda}^{(t)}
    \end{pmatrix}
  ,
      \begin{pmatrix}
       \mathbf{U}^{(t+1)} \\
       \bm{\lambda}^{(t+1)}
    \end{pmatrix}
  ,
      \begin{pmatrix}
       \mathbf{U}^{(t+2)} \\
       \bm{\lambda}^{(t+2)}
    \end{pmatrix}  
  ,\cdots
\end{align*}
where the states are traversed via the joint transition kernel

\begin{equation}
\label{eq:jointkernel}
    \big[ \bm{\lambda}^{(t+1)},\mathbf{U}^{(t+1)}\giv \bm{\lambda}^{(t)},\mathbf{U}^{(t)}\big ]  = \big[ \mathbf{U}^{(t+1)} \giv \bm{\lambda}^{(t)}\big] \big[ \bm{\lambda}^{(t+1)} \giv \bfY,\bm{\theta},\mathbf{U}^{(t+1)},\bfX\big].
\end{equation}

\begin{figure}[t!]
\captionsetup{justification=raggedright}
\centering
\begin{tikzpicture}[scale=1.0]
\begin{axis}[xmin=0,xmax=6,ymin=0,ymax=6,xlabel=$\lambda$,ylabel=$U$,xtick={0,1,2,3,4,5,6},ytick={0,1,2,3,4,5,6},xticklabels={,$\lambda^{(t)}$-2,$\lambda^{(t)}$-1,$\lambda^{(t)}$,$\lambda^{(t)}$+1,$\lambda^{(t)}$+2},yticklabels={,$\lambda^{(t)}$-2,$\lambda^{(t)}$-1,$\lambda^{(t)}$,$\lambda^{(t)}$+1,$\lambda^{(t)}$+2}
]
    \fill[fill opacity = 1.0] (3,3) circle[radius=3pt];
    \draw[draw=black, dashed, thick, fill=blue, fill opacity =0.3] (1,2) circle[radius=4pt];
    \draw[dashed, thick,fill=blue, fill opacity=0.3] (2,2) circle[radius=4pt];
    \draw[dashed, thick,fill=blue, fill opacity=0.3] (3,2) circle[radius=4pt];
    \draw[dashed, thick,fill=blue, fill opacity=0.3] (2,3) circle[radius=4pt];
    \draw[dashed, thick,fill=blue,fill opacity=0.3] (3,3) circle[radius=4pt];
    \draw[dashed, thick,fill=blue, fill opacity=0.3] (4,3) circle[radius=4pt];
    \draw[dashed, thick,fill=blue, fill opacity=0.3] (3,4) circle[radius=4pt];
    \draw[dashed, thick,fill=blue, fill opacity=0.3] (4,4) circle[radius=4pt];
    \draw[dashed, thick,fill=blue, fill opacity=0.3] (5,4) circle[radius=4pt];
    \draw[->, dashed, thick] (3,3) -- (1.1,2.1);
    \draw[->, dashed, thick] (3,3) -- (2.1,2.1);
    \draw[->, dashed, thick] (3,3) -- (3,2.1);
    \draw[->, dashed, thick] (3,3) -- (2.1,3);
    \draw[->, dashed, thick] (3,3) -- (3.9,3);
    \draw[->, dashed, thick] (3,3) -- (3,3.9);
    \draw[->, dashed, thick] (3,3) -- (3.9,3.9);
    \draw[->, dashed, thick] (3,3) -- (4.9,3.9);    
\end{axis}
\end{tikzpicture}
\caption{Representation of valid bivariate proposal states $(U^{(t+1)},\lambda^{(t+1)})$ (dashed blue circles) from state originating at $(U^{(t)},\lambda^{(t)})= (\lambda^{(t)},\lambda^{(t)})$ (solid black circle) via the bivariate kernel \eqref{eq:jointkernel} with radius $m_{\lambda}=1$ when $B=1$. Dashed directional arrows connect the origin to each proposed state. Note that $(U^{(t+1)},\lambda^{(t+1)}) = (\lambda^{(t)},\lambda^{(t)})$ is a valid proposal state.}
\label{fig:ballsamplerex}
\end{figure}
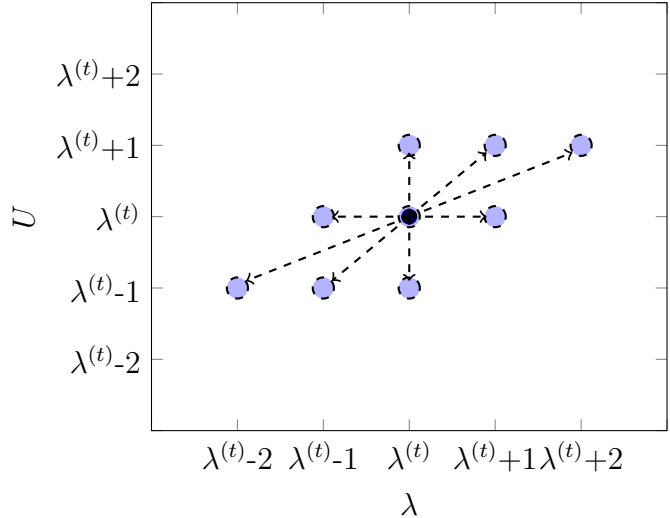
Figure \ref{fig:ballsamplerex} illustrates how \eqref{eq:jointkernel} traverses the augmented state space in the simple case where $B=1$ and $m_{\lambda}=1$.

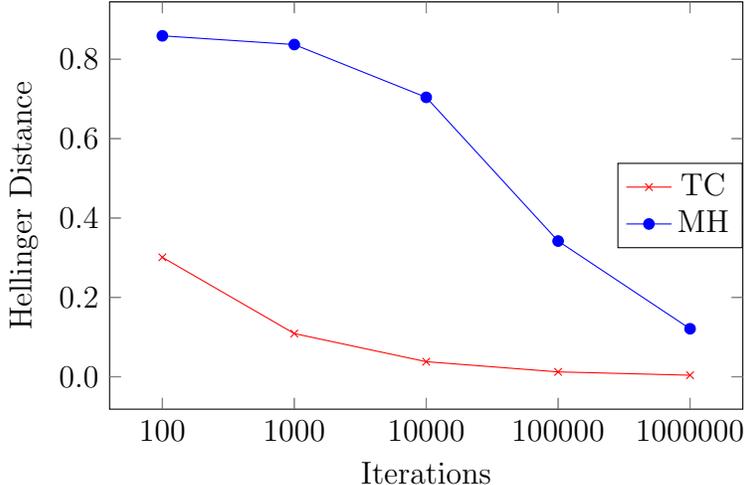
\begin{figure}
\centering
\begin{tikzpicture}
\begin{axis}[
	xlabel=Iterations,
	ylabel=Hellinger Distance,
	xticklabels={},
	yticklabels={},
    extra x ticks={1,2,...,6},
    extra y ticks={0.0,0.2,0.4,0.6,0.8,1.0},
    extra x tick labels={100,1000,10000,100000,1000000},
    extra y tick labels={0.0,0.2,0.4,0.6,0.8,1.0},
	width=10cm,height=7cm,
    legend style={at={(1,0.5)},anchor=east}]
\addplot[color=red,mark=x] coordinates {
	(1, 0.301) 
	(2, 0.1090) 
	(3, 0.0381) 
	(4, 0.0124) 
	(5, 0.0040) 
};

\addplot[color=blue,mark=*] coordinates {
	(1, 0.859) 
	(2, 0.837) 
	(3, 0.704) 
	(4, 0.342) 
	(5, 0.121) 
};

\legend{TC,MH}
\end{axis}
\end{tikzpicture}
\caption{Mean Hellinger distance (based on 100 samples) at selected iterations between estimated and target distributions under the TC and MH samplers for the example scenario presented in the Supplementary Materials section.}
\label{fig:mh_vs_ts_2}
\end{figure}

From the above construction we see the TC sampler confers certain advantages over MH proposals and direct draws from $\pi(\bm{\lambda} \giv \bfY, \bm{\theta},\bfX)$ when $\bm{\lambda}$ is high-dimensional or requires an infinite sum to obtain the normalization constant, since $q_{m_\lambda}$ is supported on a (typically) smaller set. As a result, the TC sampler provides a flexible balance between local and global search strategies via choice of $m_{\lambda}$.

As mentioned above, in the TC sampler the accept/reject mechanism is replaced by a draw from the distribution $q_{m_\lambda}$ displayed in \eqref{eq:taxicabkernel}. This strategy prevents the TC sampler from being \textit{stuck} in a certain state for long periods of time as the MH sampler might. Although $m_\lambda$ plays the role of a tuning parameter, its effects on the efficiency of the proposed sampler is far less extreme than the tuning parameter in a MH sampler. To illustrate these ideas we include in the Supplementary Materials a challenging example scenario involving an infinite but countable state space under a highly multimodal distribution. The experimental results show how the TC sampler explores the state space faster and with superior recovery of the underlying target distribution relative to a MH sampler. Figure \ref{fig:mh_vs_ts_2} summarizes this scenario, showing that the TC sampler quickly converges under the Hellinger distance whereas the MH sampler does so at a much slower rate. We include a similar comparison in terms of the total variation distance in the Supplementary Materials. 

The TC sampler also allows for sequential updating of $\bm{\lambda}$: if $\bm{\lambda}$ is organized into blocks $\bm{\lambda}_p,$ $p=1,\dots,P$, the ordered two-part Gibbs steps are now
\begin{align}
    &\mbox{1. draw } \mathbf{U}_p \sim p(\mathbf{U}_p \giv \bm{\lambda}_p) \label{eq:taxicabblockpt1} \\
    &\mbox{2. draw }\bm{\lambda}_p \sim q_{m_{\lambda}}(\bm{\lambda}_p \giv \mathbf{Y},\bm{\theta},\mathbf{U}_p,\bm{\lambda}_{-p},\mathbf{X}) \label{eq:taxicabblockpt2}
\end{align}
for all $p$, where \eqref{eq:taxicabblockpt1} depends on $Z_{m_{\lambda},p} \equiv |\mathcal{N}_{m_{\lambda}}(\bm{\lambda}_p)|$.

Finally, we note that integrating out $\mathbf{U}$ in \eqref{eq:genaugmodelfact} trivially recovers the non-augmented joint probability distribution. Further, choosing $m_{\lambda} > 0$ is necessary for ergodicity of the induced Markov chain under the TC sampler. With these results, the constructed TC sampler guarantees that the corresponding \textit{marginal} sampler on $\bm{\lambda}$ has as its stationary distribution $\pi(\bm{\lambda} \giv \mathbf{Y},\bm{\theta},\mathbf{X})$ under reasonable conditions; we refer the reader to the Supplementary Materials for the associated proofs.

\subsection{Considerations for Dimension-Changing Proposals Using the Taxicab Sampler}

The strength of the TC sampler lies in its simplicity, and it affords the ability to circumvent larger marginalization problems when performing inference on discrete model parameters; however, the sampler introduces a different set of issues when we consider more complicated models and model-fitting algorithms that involve dimension-changing proposals. In such settings we must contend with the auxiliary vector $\mathbf{U}$ injected into the model through the TC sampler, along with the fact that the model specification may lack a conjugate relationship between the likelihood function and discrete parameter prior distributions. 

We point out that the problem of dimension-changing proposals in models using the TC sampler has a different interpretation than in models using the HBS: the latter naturally allows for jumping between models encoded into binary vectors during the fitting procedure, such that the HBS proposes jumps between (potentially much) smaller- and (potentially much) larger-dimensional models with equal probability during the $\mathbf{U}$-sampling step and with no additional tools required in the proposal construction. By contrast, natural use cases for the TC sampler are more likely to involve an intrinsic ordering associated with models of differing dimension (e.g. we prefer proposing jumps in model dimension within $\pm 3$ of the current state before proposing jumps in model dimension within $\pm 10$). As such, we view dimension-jumping proposals using the TC sampler as an additional focal point worthy of discussion.

Fortunately, the construction of reasonable proposal distributions in these types of problems is mediated to a degree by the fact that we work with discrete parameter spaces instead of continuous ones, and as a result the dimension-change in these kinds of augmented chains may be handled by standard discrete-state Markov chain concepts \citep{hastie_2012}. Still, proposal construction may require more care in model frameworks where traversing models of differing dimension is non-trivial, e.g. in regression trees, and we construct one possible proposal function for tree birth and death moves within the BRT framework in Section \ref{sec:runtimeex}.

When non-conjugacy rules out the ability to efficiently marginalize out the dimension-changing parameters, we propose handling these jump proposals via a MH step. To illustrate the construction of this type of move, we consider a simple setting in which the model dimension $B$ in \eqref{eq:genaugmodelfact} is assigned a prior distribution $\pi(B)$ such that $\bm{\lambda} \sim \pi(\bm{\lambda} \giv B)$, so that our augmented joint probability model is now written as
\begin{equation}
    p(\mathbf{Y},\bm{\theta},\bm{\lambda},\mathbf{U},B \giv \mathbf{X}) = p(\mathbf{Y} \giv \bm{\theta},\bm{\lambda},B,\mathbf{X})\pi(\bm{\theta})\pi(\bm{\lambda} \giv B)\pi(B)p(\mathbf{U} \giv \bm{\lambda}),
    \label{eq:genaugdimmodelfact}
\end{equation}
where $\bm{\theta}$ may or may not depend on $B$. In this context we are interested in proposing a move from the current model state associated with $B$ to a new model state associated with dimension $B'$. Within the overall transition kernel $(\bm{\theta},\bm{\lambda},\mathbf{U},B)\rightarrow (\bm{\theta}',\bm{\lambda}',\mathbf{U}',B')$ we are primarily concerned with construction of the conditional proposal density $(\bm{\lambda},\mathbf{U} \giv B',\bm{\theta}') \rightarrow (\bm{\lambda}',\mathbf{U}' \giv B',\bm{\theta}')$ and so focus our attention on this proposal component. 

In particular, taking $(\bm{\lambda},\mathbf{U} \giv B',\bm{\theta}') \rightarrow (\bm{\lambda}',\mathbf{U}' \giv B',\bm{\theta}')$ to be \eqref{eq:jointkernel}, suitably modified to handle the change in dimension, leads to an interesting connection with the common strategy of marginalizing over continuous dimension-changing parameters in reversible-jump MCMC (RJMCMC). To generate the proposed state $(\bm{\lambda}',\mathbf{U}')$, we utilize the following order-dependent sequence of steps:
\begin{enumerate}
    \item Draw $\mathbf{a} \sim p(\mathbf{a}).$
    \item Generate $\mathbf{U}' = \delta(\bm{\lambda},\mathbf{a})$.
    \item Draw $\bm{\lambda}' \sim q_{m_{\lambda}}(\bm{\lambda}' \giv \mathbf{Y},\bm{\theta}',\mathbf{U}',B',\mathbf{X})$.
\end{enumerate}

Here $\mathbf{a}$ is a random vector of dimension $B'-B$ distributed according to $p(\mathbf{a})$ that satisfies the dimension-matching requirement of the jump proposal \citep[hereafter referred to as GR95]{green_1995}. After drawing $\mathbf{a}$, we may deterministically generate $\mathbf{U}'$ from $(\bm{\lambda},\mathbf{a})$ via a map $\delta: \Z^{B'} \rightarrow \Z^{B'}$ such that $\delta(\bm{\lambda},\mathbf{a}) = \mathbf{U}'$; we require $\delta$ be invertible in order to ensure reversibility in the resulting Markov chain. While this choice of conditional transition function depends on $\mathbf{Y}$, it leads to useful cancellations in the resulting acceptance probability calculation, which simplifies to
\begin{align}
    \alpha &\left[(\bm{\theta},\bm{\lambda},\mathbf{U},B),(\bm{\theta}',\bm{\lambda}',\mathbf{U}',B')\right] = \min\{1,A \}, \nonumber \\ 
    A= &\frac{\left[\displaystyle \sum_{\bm{\lambda}^* \in \mathcal{N}_{m_{\lambda}}(\mathbf{U}')} p(\mathbf{Y}|\bm{\theta}',\bm{\lambda}\mbox{*},B')\pi(\bm{\theta}',\bm{\lambda}\mbox{*},B')\right]p(\mathbf{U}'|\bm{\lambda}')q(B,\bm{\theta}|B',\bm{\theta}') }{\left[\displaystyle \sum_{\tilde{\bm{\lambda}} \in \mathcal{N}_{m_{\lambda}}(\mathbf{U})} p(\mathbf{Y}|\bm{\theta},\tilde{\bm{\lambda}},B)\pi(\bm{\theta},\tilde{\bm{\lambda}},B)\right]p(\mathbf{U}|\bm{\lambda})p(\mathbf{a})q(B',\bm{\theta}'|B,\bm{\theta})},
    \label{eq:dimchangemhratiosimplified}
\end{align}
and we see via \eqref{eq:dimchangemhratiosimplified} that our choice of proposal function leads to an acceptance probability ratio that weighs moving into the proposed model slice captured by $\mathcal{N}_{m_{\lambda}}(\mathbf{U}')$ from the current model slice captured by $\mathcal{N}_{m_{\lambda}}(\mathbf{U})$. It is in \eqref{eq:dimchangemhratiosimplified} that we see a correspondence with the GR95 approach of integrating out continuous dimension-changing variables in reversible-jump moves: here if we allow $m_{\lambda} \rightarrow \infty$, in the limit we obtain $\mathcal{N}_{m_{\lambda}}(\mathbf{U}) = \Z^B$ and the numerator and denominator marginal likelihood functions in \eqref{eq:dimchangemhratiosimplified} are now fully marginalized over the discrete dimension-changing vector $\bm{\lambda}$.

We do not claim this choice of transition function is optimal in some sense for a dimension-changing proposal, and for brevity also do not consider more flexible non-deterministic reverse transitions in this subsection. We refer the interested reader to \cite{brooks_2011} for an overview of strategies to construct proposal functions in RJMCMC.

\section{A Single-Tree Model for Count Data}
\label{sec:the-single-tree-model}

In this section we introduce a single-tree model for count data using a novel data distribution to serve as the likelihood function. Its advantages include easily-interpretable parameters along with the ability to handle data that displays under-, equi-, and over-dispersion. The corresponding priors encode sensible and desirable regularization and a zero-inflated extension is straightforward. Notably, this likelihood function is not in the exponential family and the selected prior distributions are not conjugate.

To place our proposed model in context, we begin with brief reviews of Bayesian regression tree models and models designed to handle count data in Subsections \ref{ss:brt-review} and \ref{ss:count-models} respectively. We introduce our proposed model in Subsection \ref{ss:single-tree-likelihood}.

\subsection{A Review of Bayesian Regression Tree Models}
\label{ss:brt-review}

\subsubsection{Regression Tree Models}

\begin{figure}[t!]
\captionsetup{justification=raggedright}
\centering
\begin{tikzpicture}[scale=0.7,every tree
node/.style={draw,circle,minimum size=2cm},
   level distance=3cm,sibling distance=2cm, 
   edge from parent path={(\tikzparentnode) -- (\tikzchildnode)}]
\Tree [.\node [red,label=below:$\eta_1$] {$X_{1} < 4$}; 
    \edge[red]; 
    [.\node[red,label=below:$\eta_2$]{$X_{2} < 2$}; \node [label=below:$\eta_4$] {$\lambda_4=4$}; \edge[red]; \node [red,label=below:$\eta_5$] {$\lambda_5=3$};
    ]
    \node [label=below:$\eta_3$] {$\lambda_3=2$}; 
    ]
\end{tikzpicture}
\qquad
\begin{tikzpicture}[scale=0.7]
\begin{axis}[xmin=0,xmax=6,ymin=0,ymax=6,xlabel=$X_1$,ylabel=$X_2$,samples=50]
    \draw[black, ultra thin, dashed] (4,0) -- (4,6);
    \draw[black, ultra thin, dashed] (0,2) -- (4,2);
    \node at (axis cs: 2,1) {$\lambda_4=4$};
    \node at (axis cs: 2,4) {$\lambda_5=3$};
    \node at (axis cs: 5,3) {$\lambda_3=2$};
\end{axis}
\end{tikzpicture}

\caption{Left: An example tree $(\mathcal{T},\mathbf{M})$ containing two internal nodes $(\eta_1,\eta_2)$ with rules and three terminal nodes $(\eta_3,\eta_4,\eta_5)$ with associated parameters. The example vector $(x_1,x_2) = (3,3)$ would be sorted along the path highlighted by the red nodes and collected in $\eta_5$; the corresponding path from root node to $\eta_5$ is represented by $\X_5 = \{x_1 < 4 \cap x_2 \geq 2 \}$. Right: an equivalent representation of $(\mathcal{T},\mathbf{M})$ demonstrating how the tree structure partitions the covariate space $(X_1,X_2)$ into hyperrectangles and assigns parameter values to each. Here $\X_5$ corresponds to the top-left hyperrectangle with parameter value $\lambda_5 = 3$.}
\label{fig:treeex1}
\end{figure}
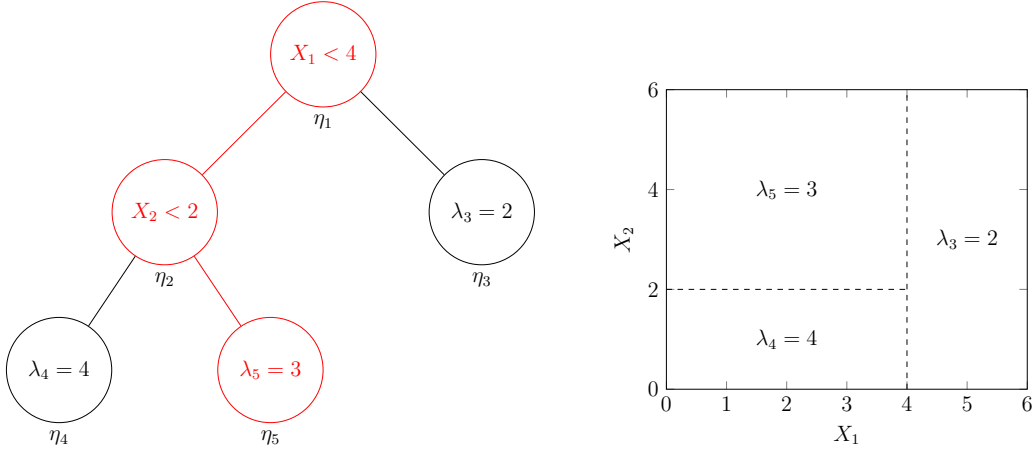

Each binary decision tree $\mathcal{T}$ is comprised of vertices (which we refer to as nodes or $\eta$'s), edges, and splitting rules. Nodes positioned at the terminus of a tree branch are labelled {\em terminal} nodes and equipped with parameters that describe the data associated with them; non-terminal nodes are labelled 
{\em internal} nodes and equipped with binary rules. An internal node $\eta$ may also be referred to as a {\em parent} node; the connected nodes below it are referred to as left and right {\em child} nodes based on their relative position to the parent. The left panel of Figure \ref{fig:treeex1} visualizes an example tree $\mathcal{T}$ with parent and child nodes labeled.

A regression tree $\mathcal{T}$ partitions the covariate space according to the binary internal node rules that comprise its interior. These internal node rules are often equivalently referred to as {\em decision} or {\em splitting} rules and may be compactly described using the notation $(v,c_v)$, where $v \in \{1,...,p \}$ corresponds to the $v$th dimension of the covariate space, and $c_v \in \mathcal{C}_v$ corresponds to a comparison value or set in the $v$th dimension, with the relation $(v,c_v) = \{x_{iv} < c_v \}$ or $(v,c_v) = \{x_{iv} \in c_v \}$. 

In tree models, $\vert\mathcal{C}_v\vert$ is finite since each covariate dimension is typically discretized over an equally-spaced grid according to some resolution in order to ensure the tree space does not become prohibitively difficult to explore. This grid resolution is specified via a user-selected number of ``cutpoints" $\zeta_v$ for any numeric covariate $\mathbf{x}_v$, and otherwise $\zeta_v = \text{dim}(\mathbf{x}_v)$ for categorical covariates \citep[however other encodings are possible\textemdash see for instance][]{dorogush_2018}. The discretization is based on the observed range of covariate values so that in the $v$th dimension, each cutpoint $c_v$ is such that $c_v \in (\min_i({x_{iv}}), \max_i({x_{iv}}))$ if $\mathbf{x}_v$ is numeric, or else a subset of categories $c_v \subset \mathbf{x}_v$ if $\mathbf{x}_v$ is categorical. 

Specifically, a decision rule $(v,c_v)$ operates on a continuous covariate vector $\mathbf{X}_i$ as follows. The internal node rule $(v,c_v)$ evaluates the event $\{ x_{iv} < c_v \}$: if $\{x_{iv} < c_v\}$ is true, $\mathbf{X}_i$ is deterministically assigned to the left child node; otherwise $\mathbf{x}_i$ is assigned to the right child node. Each $\mathbf{X}_i$ is sorted along a resultant path of internal nodes in this fashion until a terminal node $\eta$ is reached, at which point it is collected in $\eta$. To each terminal node we also assign a value $\lambda_b$ where $b=1,\ldots, B$ indexes a terminal node. In this framework, the tree structure defines a function $g:\R^p \to \R$. If covariate $\mathbf{X}_i$ is assigned to terminal node $\eta_b$ then we set 
$$
g(\mathbf{X}_i) = \lambda_b .
$$
In a probabilistic tree model, $\mathcal{T}$ may be used to flexibly describe statistical models for data $(Y_i,\mathbf{X}_i)$. Using the notation described above, let $g(\cdot)$ denote the function described by the tree $\mathcal{T}$. The values $\mathbf{M} = (\lambda_1, \ldots, \lambda_B)$ assigned to the terminal nodes in $\mathcal{T}$ are viewed as parameters and $g(\cdot)$ is the regression function. This model is summarized as 
\begin{eqnarray}
\begin{aligned}
\label{eq:treemeanfun}
    \bfY \giv \bfX, \mathbf{M}, \bftheta &\sim f(\cdot \giv \bfX, \mathbf{M}, \bftheta)\ ,\quad \mbox{where}\\
    \E(Y_i \giv \mathbf{X}_i, \mathbf{M}) &= g(\mathbf{X}_i) = \sum_{b=1}^B \lambda_b \I(\mathbf{X}_i \in \X_b), 
\end{aligned}
\end{eqnarray}
where $f$ is a generic notation for a likelihood function selected by the user and $\X_b$ represents the intersection of the regions described by the internal nodes in $\mathcal{T}$ that form the path from the root node to terminal node $\eta_b$. The right panel of Figure \ref{fig:treeex1} shows how such a tree $(\mathcal{T},\mathbf{M})$ partitions and describes the covariate space according to \eqref{eq:treemeanfun}.
Further assumptions (conditional independence, for example) will allow us to write the model \eqref{eq:treemeanfun} in a simplified, more manageable form, as seen in the sections to come. In addition to the parameter vector $\bm{\lambda}$ that is used to describe the mean function, we allow for other parameters, collected in $\bftheta$, to be a part of the probabilistic model \eqref{eq:treemeanfun}. Such parameters may model other components of the model, such as the variance if using a Gaussian likelihood. 

\subsubsection{Bayesian CART}

\cite{cgm_1998} (hereafter referred to as CGM98) introduce the Bayesian analogue of the CART model along with the MH analogue of \cite{breiman_1984}'s CART algorithm for fitting tree models.

\paragraph{Likelihood Function and Prior Distribution Specification in Bayesian CART}

In the Bayesian CART model, the joint prior distribution $\pi(\mathcal{T},\mathbf{M},\bm{\theta})$ is assumed to factorize as
\begin{align}
    \pi(\mathcal{T},\mathbf{M},\bm{\theta}) &= \pi(\mathbf{M} \giv \mathcal{T})\pi(\mathcal{T})\pi(\bm{\theta}), \label{eq:cgmpriorfact1} \\
    \pi(\mathbf{M} \giv \mathcal{T}) &= \prod_{b=1}^B \pi(\lambda_b \giv \mathcal{T}), \label{eq:cgmpriorfact2}
\end{align}
for the terminal node parameters $\lambda_b$, allowing for a prior distribution on the tree topology $\mathcal{T}$ that does not depend on $\mathbf{M}$ and codifying the assumption of a priori independence of terminal node parameters given $\mathcal{T}$. Typically, conjugate prior distributions are assigned to the $\lambda_b$'s and $\bm{\theta}$.

Notably, CGM98 specify $\pi(\mathcal{T})$ implicitly. Tree complexity is controlled via the prior probability that a node $\eta$ will be internal/non-terminal, defined as 
\begin{align*}
\pi(\eta \text{ is internal}) &= \alpha(1+d(\eta))^{-\beta},\ \alpha \in (0,1),\ \beta >0, \\
\pi(\eta \text{ is terminal}) &= 1- \pi(\eta \text{ is internal}).
\end{align*}

Each splitting rule $(v,c_v)$ is determined by the choice of available covariates $\mathbf{x}_v$, where $v$ is selected uniformly among all available predictors, and the choice of cutpoint value (or category) $c_v$ is also selected uniformly among the available generated cutpoints given $v$ and $\mathcal{T} \setminus \eta$.

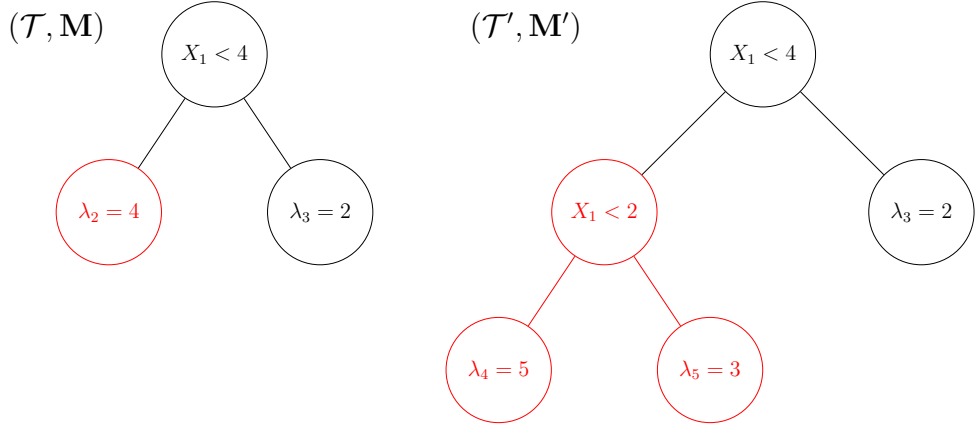
\begin{figure}[t!]
\centering
\begin{tikzpicture}[scale=0.7,anchor=center,baseline,every tree
node/.style={draw,circle,minimum size=2cm},
   level distance=3cm,sibling distance=2cm, 
   edge from parent path={(\tikzparentnode) -- (\tikzchildnode)}]
\Tree [.\node (root) {$X_{1} < 4$}; 
    \node[red]{$\lambda_2=4$}; \node {$\lambda_3=2$}; 
 ]
 \node[align=center] at (-3,0.5) {$(\mathcal{T},\mathbf{M})$};
\end{tikzpicture}
\qquad
\begin{tikzpicture}[scale=0.7,anchor=center,baseline,every tree
node/.style={draw,circle,minimum size=2cm},
   level distance=3cm,sibling distance=2cm, 
   edge from parent path={(\tikzparentnode) -- (\tikzchildnode)}]
\Tree [.{$X_{1} < 4$} 
    [.\node[red]{$X_{1} < 2$};
       \edge[red]; [.\node[red]{$\lambda_4=5$}; ] \edge[red]; [.\node[red]{$\lambda_5=3$}; ]
    ]
    [.{$\lambda_3=2$} ]
 ]
  \node[align=center] at (-4.5,0.5) {$(\mathcal{T}',\mathbf{M}')$};
\end{tikzpicture}
\caption{Left: a tree $(\mathcal{T},\mathbf{M})$ with red-highlighted terminal node selected for birth. Right: an updated tree $(\mathcal{T}',\mathbf{M}')$ after birth, with the updated tree structure again highlighted in red. Here the node $\eta_2$ has been converted to an internal node with rule $X_1 < 2$ and assigned two child terminal nodes. Note that the choice of new rule leads to a coherent re-partitioning of the covariate space. The reverse death move is represented by the transformation $(\mathcal{T}',\mathbf{M}') \rightarrow (\mathcal{T},\mathbf{M})$.}
\label{fig:treebdex}
\end{figure}

\paragraph{Fitting the Bayesian CART Model}

Fitting a Bayesian CART model requires mechanisms for exploring the conditional posterior distributions for $\mathcal{T}$, $\mathbf{M}$, and $\bm{\theta}$. Conveniently, the choice of conjugate prior distributions for the terminal node parameters collected in $\mathbf{M}$ as well as $\bm{\theta}$ results in straightforward Gibbs updates for these model parameters.

CGM98 propose searching over the tree space via the use of a top-down stochastic tree-generating process. Broadly, trees (beginning with a single root node) are grown by first proposing a {\em birth} move (selected with probability $\pi(\text{birth})$) and then uniformly selecting a terminal node to be split; tree pruning occurs via {\em death} moves (selected with probability $\pi(\text{death})$) performed on a uniformly-selected next-to-terminal node. These moves are illustrated in Figure \ref{fig:treebdex}. A successful birth move converts the selected terminal node $\eta$ into an internal node, which is then assigned a splitting rule and left and right child (terminal) nodes. A successful death move reverses this birth process. Exploration of the posterior tree space in the CGM98 model formulation is also facilitated by two additional proposals, swap and change; proposals affecting $\mathcal{T}$ are handled via MH steps after integrating out the dimension-changing mean vector $\mathbf{M}$, made possible due to the choice of conjugate prior distributions on the $\lambda_b$'s.

\subsection{Count Models}
\label{ss:count-models}

For count models, typical likelihood choices include the Poisson and negative binomial distributions, with the latter often desirable due to its ability to model overdispersion. Other likelihood choices for count data models have been introduced that allow for even more flexible modeling of mean-variance relationships (i.e. underdispersion), including the COM-Poisson \citep{conway_1962} and double Poisson \citep{efron_1986}, but these have historically suffered from computational and parameter interpretation issues that have prevented more widespread adoption \citep{sellers_2010}.

In the tree literature, \cite{murray_2020} extends the BART framework to accommodate multinomial logistic and count regression models, utilizing the sum-of-trees approach to build up appropriate transformations of mean functions of interest. Key to their model is a sampling algorithm that allows for blocked MCMC updating of each tree $\mathcal{T}_h,$ $h=1,...,m$, and its parameters $\mathbf{M}_h$ while holding $(\bm{\mathcal{T}}_{-h},\mathbf{M}_{-h})$ fixed. In the case of count data, Murray achieves this by augmenting a Poisson or negative binomial likelihood with additional latent variables to allow for tractable integrated marginal likelihoods and introduces new conjugate prior distributions. The resulting closed-form marginal likelihoods are thus available to update individual trees via MH, and  updating each tree's terminal node parameters may be performed using Gibbs steps after latent variable updates, yielding a non-backfitting update algorithm that still resembles BART's fast-update scheme; however, a key disadvantage of Murray's count model formulation is the number of data-dependent latent variable updates required when modeling zero-inflation and overdispersion.

\subsection{The Likelihood Function}
\label{ss:single-tree-likelihood}

\begin{figure}[ht]
    \centering
    \begin{subfigure}[b]{0.8\linewidth}
    \includegraphics[width=\linewidth]{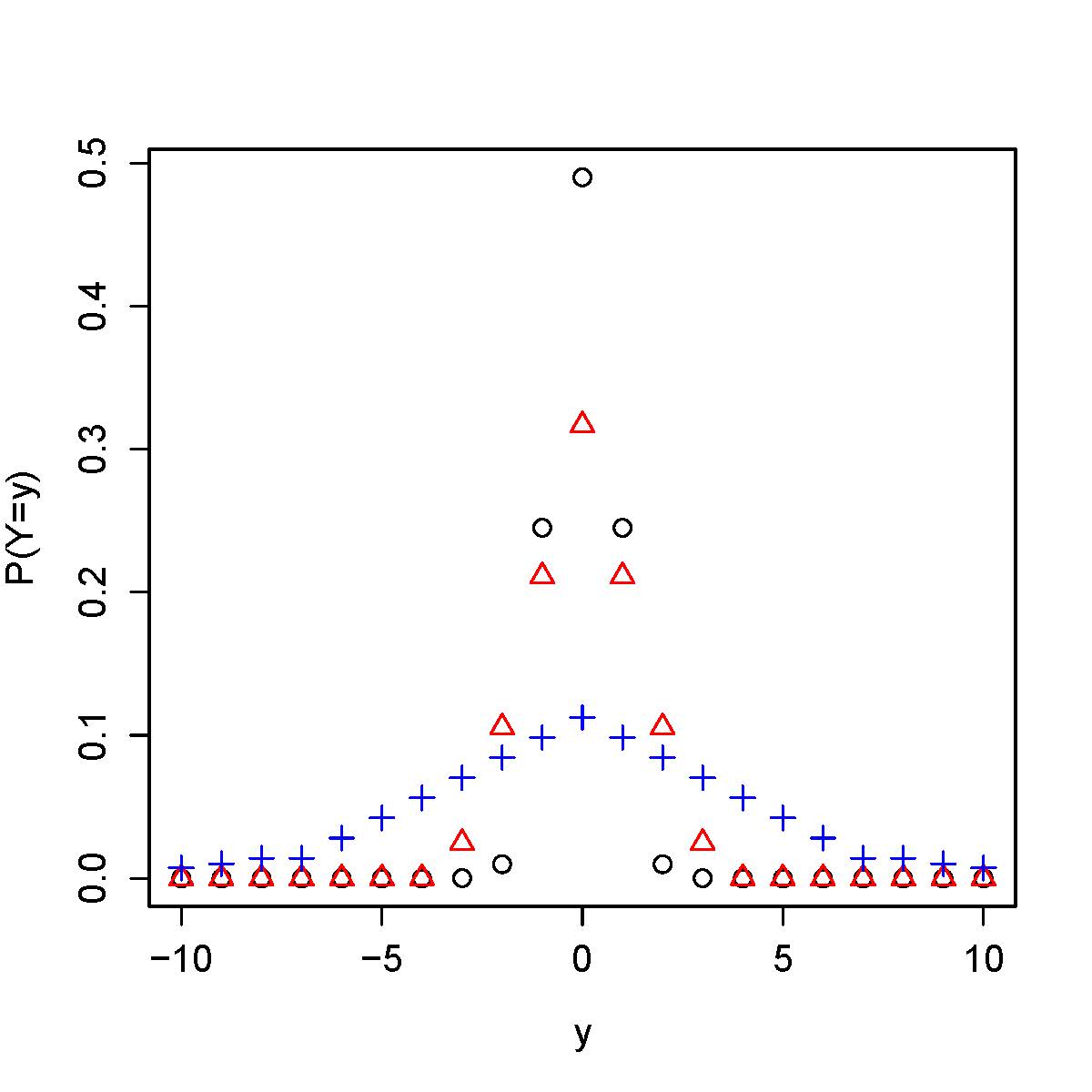}
    \end{subfigure}%
    \caption{Several examples of the tent distribution using the tent and tail definitions given in (\ref{eq:tentpmf}) and (\ref{eq:tentpmftail}). $P_{0.01}(0,1)$ is represented by the black circles, $P_{0.025}(0,2)$ is represented by the red triangles, and $P_{0.05}(0,7)$ is represented by the blue crosses.}
    \label{fig:tentpmfex}
\end{figure}

Our proposed likelihood function is fully specified by three parameters: a location parameter $\lambda \in \Z_{\geq 0}$; a scale parameter $k \in \Z$; and a tail mass parameter $t \in [0,0.5)$.  The tail mass parameter controls the probability assigned to each tail of the resulting distribution. The definition of this distribution follows a piece-wise construction and is written as follows:
\begin{numcases}{p_t(Y \giv \lambda,k) =}
      (1-2t) \frac{k+1-|Y-\lambda|}{(k+1)^2}, &  $|Y-\lambda| \leq k$ \label{eq:tentpmf} \\
      t p^{*} (1-p^{*})^{|Y-\lambda|-k-1} & $|Y-\lambda| > k$, \label{eq:tentpmftail}
\end{numcases}
where we define 
$$
p^{*} \equiv \min \left\{0.99,\frac{1-2t}{t(k + 1)^2} \right\}
$$
This distribution, which we term a {\em tent} probability mass function (pmf), is unimodal and symmetric about $\lambda$. The parameter $k$ controls the range of $Y$ values centered about $\lambda$ in the distribution, in that the range of values contained in the middle $100(1-2t)\%$ of the distribution (the ``tent") is always $2k$. The tails follow a geometric distribution and each contain $t$ total mass. Figure \ref{fig:tentpmfex} visualizes this distribution for several settings of  $\lambda$, $k$, and $t$.
We treat $t$ as a hyperparameter to be fixed prior to the outset of any modeling problem. The choice of including separate scale and tail parameters in this pmf allows for additional flexibility in modeling data: the tail parameter $t$ may be calibrated to describe relevant features of interest at the outset of the problem, whereas the behavior of data closer to the mode can be described separately using $k$. Moving forward, we refer to the PMF parameterized by $(\lambda,k,t)$ using $P_t(\lambda,k)$. Probabilities according to $Y \sim P_t(\lambda,k)$ will be referred to using $p_t(Y|\lambda,k)$, with the conditioning suppressed in cases where the parameterization is clear.

\subsection{Prior Distributions on $\lambda$ and $k$}

Since we utilize a single-tree approach for our proposed model, we maintain the implicit prior formulation on $\mathcal{T}$ as specified in CGM98, along with the prior factorization assumptions in (\ref{eq:cgmpriorfact1}) and (\ref{eq:cgmpriorfact2}). For simplicity, given the tree $\mathcal{T}$ we define a discrete uniform prior distribution on the terminal node mean parameters, i.e.
\begin{equation}
    \lambda_b \giv \mathcal{T},d_1,d_2 \stackrel{ind.}{\sim} \text{DU}\{d_1,d_2\},\ d_1,d_2 \in \mathbb{N}_{\geq 0},\ d_1 < d_2,
\end{equation} 
in effect allowing the likelihood to guide our search algorithm toward good choices of $\lambda_b$. In application, we have taken $d_1 = \min(y)$ and $d_2 = \max(y)$ as reasonable but vague default hyperparameter values.

We propose using a similar function to the one defined in \eqref{eq:tentpmf} and \eqref{eq:tentpmftail} as the prior distribution on each $k_i$. Since the tent pmf places positive mass on $\Z$, we adjust the likelihood function accordingly to ensure appropriate behavior of the likelihood scale parameter; in particular, we work with the exponentiated likelihood parameterization
\begin{equation}
\label{eq:finalpmflikelihood}
    Y_i \giv \lambda_i,k_i,t \simind P_t(\lambda_i,\floor{e^{k_i}}), \qquad i=1,...,n,
\end{equation}
so that $\floor{e^{k_i}} \in \Z_{\geq 0}$ as previously required. Moving forward, we will use (\ref{eq:finalpmflikelihood}) in all instances involving the likelihood function unless otherwise noted. We specify the prior model on $k_b$ associated to terminal node $\eta_b$ as
\begin{equation}
\label{eq:scalepriorinfsupportfull}
k_b \giv \lambda_b,\kappa,\beta_k,t_k,\mathcal{T} \simind P_{t_k} \left( \floor*{\frac{\kappa }{2^{d(\eta_b)} } }, \floor*{ \frac{\widetilde{\log}( \lambda_b)}{(1+d(\eta_b))^{\beta_k}} } \right),\ \kappa \in \Z_{\geq 0},\ \beta_k \geq 0,
\end{equation}
where
\begin{equation*}
\widetilde{\log}(\lambda)  =
    \begin{cases}
      0, &  \lambda \leq 1, \\
      \log(\lambda), & \lambda > 1.
\end{cases}
\end{equation*}

The prior mode in (\ref{eq:scalepriorinfsupportfull}) is a function of a hyperparameter value $\kappa$, which may be chosen to reflect some belief in the true underlying degree of dispersion if prior information is available, or otherwise chosen to provide an initial guide for the likelihood fit at the grand mean model level. The mode is also a function of node depth with respect to its terminal node $\eta_b$ and codifies the belief that terminal nodes at deeper levels of a tree, reflecting more complex regions of the response surface, should accordingly seek to explain less variability in the data.

The prior scale also encodes desirable behavior. The denominator $(1+d(\eta_b))^{\beta_k}$ utilizes the node depth penalization function in the CGM98 tree prior, where in this setting $\beta_k$ controls the rate at which the prior dispersion increases or decreases; allowing $\beta_k \rightarrow \infty$ has the effect of concentrating the scale prior on its current location parameter, while letting $\beta_k \rightarrow 0$ allows the current $\lambda_b$ to fully dictate the dispersion of the scale prior distribution.

From above, the full hierarchical model is specified as follows:
\begin{equation}
\begin{aligned}
Y_i \giv t,\mathcal{T},\lambda_i,k_i,X_i &\simind P_t(\lambda_i,\floor*{\exp\{k_i \}}),  \\
\lambda_b \giv d_1,d_2,\mathcal{T} &\simiid \text{DU}\{d_1,\dots,d_2 \},\ b=1,\dots,B \\
k_b \giv \lambda_b,\kappa,\beta_k,t_k,\mathcal{T} &\simind P_{t_k} \left( \floor*{\frac{\kappa }{2^{d(\eta_b)} } }, \floor*{ \frac{\widetilde{\log}( \lambda_b)}{(1+d(\eta_b))^{\beta_k}} } \right),\ b=1,\dots,B \\
\mathcal{T} \giv \alpha,\beta,\zeta_v &\sim \pi(\mathcal{T}|\alpha,\beta,\zeta_v).
\end{aligned}
\end{equation}

\section{Comparing the Algorithms for the Single-Tree Count Data Example}
\label{sec:runtimeex}

In this example, we generate observations from a true underlying model and attempt to recover the correct tree structure and terminal node parameters via our single-tree count model (see Section \ref{sec:the-single-tree-model}) using the taxicab update algorithm, and compare to a na{\"i}ve MH algorithm in order to illustrate the gains in computational speed under the proposed approach. We also considered other example settings, including data exhibiting excess zeroes, for which we developed a ZI extension to our TC-augmented model. These results are available in the Supplementary Materials and were similar to what is reported here. In the present example we considered the following two-covariate true data-generating model:
\begin{equation}
    Y_i \giv \bfX_{i} \stackrel{ind.}{\sim} P_t(g(\bfX_i),\floor{e^k}=\floor{e^2}), \label{eq:runtimeextruemodel}
\end{equation}
where
\begin{equation}
\label{eq:tentpmfexp1lambdafun}
 g(\bfX_i) = \begin{cases} 
      10 , &  x_{i1} \leq 5, x_{i2} \leq 5 \\
      20 , &  x_{i1} \leq 5, x_{i2} > 5 \\
      30 , &  x_{i1} > 5, x_{i2} \leq 5 \\
      40 , &  x_{i1} > 5, x_{i2} > 5,
   \end{cases}
\end{equation}
with $X_{ij} \sim \text{Unif}(0,10)$ for $i = 1,...,n$ and $j=1,2$. For simplicity we took $t=0$ in this true model so that \eqref{eq:runtimeextruemodel} places positive probability on a finite set of values.

We proceeded to generate $n=10,\ 100,$ and $1000$ observations according to this true model for use in our fitting procedures. 

\subsection{Setup}
\label{ss:runtimesetup}

In the non-ZI model, we begin by injecting auxiliary vectors $\mathbf{U} = (U_1,...,U_B) \in \Z^B$ and $\mathbf{R} = (R_1,...,R_B) \in \Z^B$ into our joint distribution, so that we now work with an augmented joint distribution $p(\bfY,\mathcal{T},\bm{\lambda},\bm{k},\mathbf{U},\mathbf{R})$. We note in particular that the auxiliary vectors in this model are only of dimension $B$, whereas in other approaches they are typically of dimension $n$ \citep[e.g.][]{albert_1993,murray_2020}. We further require the ability to factorize this joint distribution as
\begin{equation}
    p(\bfY,\mathcal{T},\bm\lambda,\bm k,\mathbf{U},\mathbf{R}) 
    = p(\bfY,\mathcal{T},\bm\lambda,\bm k)p(\mathbf{U} \giv \bm \lambda)p(\mathbf{R} \giv \bm k), \label{eq:augjointdistn}
\end{equation}
where
\begin{align}
p(\bfY,\mathcal{T},\bm\lambda,\bm k) &= \left\{ \prod_{b=1}^B \Lik(\lambda_b,k_b \giv \bfY,\cdot) \pi(\lambda_b \giv \mathcal{T}) \pi(k_b \giv \lambda_b,\mathcal{T}) \right\} \pi(\mathcal{T}), \label{eq:augjointdistn1} \\
p(\mathbf{U} \giv \bm \lambda)p(\mathbf{R} \giv \bm k) &= \prod_{b=1}^B p(U_b \giv \lambda_b) p(R_b \giv k_b). \nonumber
\end{align}

Utilizing $B$ total blocks for updating both $\bm{\lambda}$ and $\bm{k}$, the complete TC sampler in this problem is now constructed as
\begin{align}
         U_b^{'} & \sim p(U_b^{'} \giv \lambda_b), \nonumber \\ 
         \lambda_b^{'} & \sim
         q_{m_{\lambda}}(\lambda_b^{'} \giv \bfY,\mathcal{T},k_b,U_b^{'}) \propto p(\bfY,\mathcal{T},\lambda_b^{'},k_b)\mathbb{I}\{\lambda_b^{'} \in \mathcal{N}_{m_{\lambda}}(U_b^{'})\}, \label{eq:balllambdaq} \\
         R_b^{'} & \sim p(R_b^{'} \giv k_b), \nonumber \\
         k_b^{'} & \sim q_{m_k}(k_b^{'} \giv \bfY,\mathcal{T},\lambda_b,R_b^{'}) \propto p(\bfY,\mathcal{T},\lambda_b^{'},k_b^{'})\mathbb{I}\{k_b^{'} \in \mathcal{N}_{{m_k}}(R_b^{'}) \}, \label{eq:ballkq} 
\end{align}
for $b=1,\ldots,B.$ Note that several cancellations follow in the numerator and denominator of both (\ref{eq:balllambdaq}) and (\ref{eq:ballkq}) according to (\ref{eq:augjointdistn1}), including $\pi(\mathcal{T})$, $\pi(\lambda_b \giv \mathcal{T})$, and all likelihood contributions from terminal nodes besides $\eta_b$; the auxiliary distribution for the fixed terminal node parameter also naturally cancels from these kernels ($p(R_b \giv k_b)$ for the update of $\lambda_b$, $p(U_b^{'} \giv \lambda_b^{'})$ for the update of $k_b$).

For the na{\"i}ve MH algorithm, we utilized simple MH moves with discrete uniform proposal distributions to update $(\lambda_b,k_b)$ values, along with tree birth and death moves using the marginal likelihood approach described in CGM98. In both search approaches we utilized birth and death moves along with \cite{pratola_2016}'s cutpoint perturb proposal as an added way to facilitate exploration of the posterior tree space.

Here the tree birth and death moves require accounting for a change of dimension in both $\bm{\lambda}$ and $\bm{k}$. Since the repertoire of tree moves we work with are local perturbations by design, we proceed to describe the construction of a dimension-changing move for tree birth proposals with respect to the local change about a terminal node $\eta_b$ that has been selected for birth, with its proposed left and right child nodes denoted by $\eta_{b(l)}$ and $\eta_{b(r)}$ respectively and the proposed rule $(v',c')$ assigned to $\eta_b$ in the new structure $\mathcal{T}'$. Further let $\tilde{\bm{\lambda}} = (\lambda_1,\dots,\lambda_{b-1},\lambda_{b(l)},\lambda_{b(r)},\lambda_{b+1},\dots,\lambda_B)$, with equivalent definitions for $\tilde{\bm{k}}$, $\tilde{\mathbf{U}}$, and $\tilde{\mathbf{R}}$. In this joint setting we also let $\bm m = (m_{\lambda},m_k)$. Conveniently, the transition proposals between $\mathcal{T}$ and $\mathcal{T}'$ in our model are still handled as in CGM98, and so the immediate problem is that of the conditional transitions
\begin{align}
    \lambda_b &\rightarrow (U_{b(l)},U_{b(r)}) \label{eq:lambdat1} \\
    (U_{b(l)},U_{b(r)}) &\rightarrow (\lambda_{b(l)},\lambda_{b(r)}) \label{eq:lambdat2} \\
    k_b &\rightarrow (R_{b(l)},R_{b(r)}) \label{eq:kt1} \\
    (R_{b(l)},R_{b(r)}) &\rightarrow (k_{b(l)},k_{b(l)}) \label{eq:kt2}
\end{align}
given $\mathcal{T}$ and $\mathcal{T}'$. 

Per GR95, one useful way to resolve the transitions \eqref{eq:lambdat1} and \eqref{eq:kt1} is through dimension-matching. Specifically, we generate discrete random variables $a_{\lambda}$ and $a_k$ and subsequently define an invertible and deterministic mapping $\delta: \Z^2 \rightarrow \Z^2$ to match the dimension of our current state to that of the proposed state: for some $\theta,a \in \Z$ we define
\begin{align}
    \delta(\theta,a) &= \left(\delta_1[\theta,a],\delta_2[\theta,a]\right) \\
    &= \left(\theta - \floor*{\frac{a}{2}}, \theta + \ceil*{\frac{a}{2}}\right),
\end{align}
with inverse $\delta^{-1}: \Z^2 \rightarrow \Z^2$ given by
\begin{align}
    \delta^{-1}(x,y) &= \left(\delta^{-1}_1[x,y],\delta^{-1}_2[x,y]\right) \\
    &= \left(\floor*{\frac{x+y}{2}} ,y-x\right),\ x,y \in \Z,
\end{align}

so that we obtain the desired transitions in this problem via the following sequence of operations:

\begin{enumerate}
    \item Generate random scalars $a_{\lambda} \sim \text{DU}\{-2m_{\lambda},2m_{\lambda}\},\ a_k \sim \text{DU}\{-2m_k,2m_k \}$.
    \item Given $a_{\lambda}$, generate the auxiliary variables $(u_{b(l)},u_{b(r)}) = \delta(\lambda_b,a_{\lambda}) = \left(\lambda_b-\floor*{\frac{a_{\lambda}}{2}},\lambda_b+\ceil*{\frac{a_{\lambda}}{2} }\right)$.
    \item Given $a_k$, generate the auxiliary variables $(r_{b(l)},r_{b(r)}) = \delta(k_b,a_{k}) = \left(k_b-\floor*{\frac{a_{k}}{2}},k_b+\ceil*{\frac{a_k}{2}}\right)$.
\end{enumerate}

Note that the choice of support for the random scalars $a_{\lambda}$ and $a_k$, in tandem with the definition of the function $\delta$, ensures that the newly-proposed auxiliary variables in steps \#2 and \#3 of the above operation are contained respectively within the neighborhoods centered at the current parameter values $\lambda_b$ and $k_b$ in birth proposals, as required under the TC sampler framework.

From here the transitions \eqref{eq:lambdat2} and \eqref{eq:kt2} are proposed according to
\begin{align}
    (\lambda_{b(l)},k_{b(l)}) &\sim q_{\bm m}(\lambda_{b(l)},k_{b(l)} \giv \mathbf{Y},\mathcal{T}',U_{b(l)},R_{b(l)}), \label{eq:ljointpropkernel1} \\
    (\lambda_{b(r)},k_{b(r)}) &\sim q_{\bm m}(\lambda_{b(r)},k_{b(r)} \giv \mathbf{Y},\mathcal{T}',U_{b(r)},R_{b(r)}), \label{eq:rjointpropkernel1}
\end{align}
where (\ref{eq:ljointpropkernel1}) and (\ref{eq:rjointpropkernel1}) follow the general definition
\begin{equation}
\begin{aligned}
    q_{\bm m}(\lambda_{b},k_{b} \giv \mathbf{Y},\mathcal{T},U_{b},R_{b}) &\propto p(\mathbf{Y},\mathcal{T},\lambda_{b},k_{b})\mathbb{I}\{\lambda_{b},k_{b} \in \mathcal{N}^2_m(U_{b},R_{b}) \}, \\
    \mathbb{I}\{\lambda_{b},k_{b} \in \mathcal{N}^2_m(U_{b},R_{b}) \} &= \mathbb{I}\{(\lambda_{b},k_b) \in \mathcal{N}_{m_{\lambda}}(U_{b}) \times \mathcal{N}_{m_k}(R_{b}) \}. \nonumber
\end{aligned}
\end{equation}
The full reverse transition is then generated according to
\begin{align}
    U_b &= \delta_1^{-1}(\lambda_{b(l)},\lambda_{b(r)}) = \floor*{ \frac{1}{2}(\lambda_{b(l)}+\lambda_{b(r)}) }, \label{eq:bproprevmove11} \\
    R_b &= \delta_1^{-1}(k_{b(l)},k_{b(r)}) = \floor*{ \frac{1}{2}(k_{b(l)}+k_{b(r)}) }, \label{eq:bproprevmove22} \\
    (\lambda_{b},k_{b}) &\sim q_{\bm m}(\lambda_{b},k_{b} \giv \mathbf{Y},\mathcal{T},U_{b},R_{b}), \label{eq:bproporevmove33}
\end{align}
where \eqref{eq:bproprevmove11} and \eqref{eq:bproprevmove22} again allow for a deterministic mapping from the proposed higher-dimensional model to the lower-dimensional one, and \eqref{eq:bproporevmove33} simply calculates the joint probability of the reverse transitions $(\lambda_{b(l)},\lambda_{b(r)}) \rightarrow (\lambda_b,U_b)$ and $(k_{b(l)},k_{b(r)}) \rightarrow (k_b,R_b)$.

The acceptance probability for this birth move is calculated as
\begin{align}
    &\alpha [(\mathcal{T},\bm{\lambda}, \bm{k},\mathbf{U},\mathbf{R}),(\mathcal{T}',\tilde{\bm{\lambda}},\tilde{\bm{k}},\tilde{\mathbf{U}},\tilde{\mathbf{R}})] \label{eq:mhratio1} \\
    = &\text{min}\left\{1, \frac{ \pi(\mathcal{T}',\tilde{\bm{\lambda}},\tilde{\bm{k}},\tilde{\mathbf{U}},\tilde{\mathbf{R}} \giv \mathbf{Y},\cdot) q(\mathcal{T},\bm{\lambda}, \bm{k},\mathbf{U},\mathbf{R} \giv \mathcal{T}',\tilde{\bm{\lambda}},\tilde{\bm{k}},\tilde{\mathbf{U}},\tilde{\mathbf{R}}) }{ \pi(\mathcal{T},\bm{\lambda}, \bm{k},\mathbf{U},\mathbf{R} \giv \mathbf{Y},\cdot)q(\mathcal{T}',\tilde{\bm{\lambda}},\tilde{\bm{k}},\tilde{\mathbf{U}},\tilde{\mathbf{R}} \giv \mathcal{T},\bm{\lambda}, \bm{k},\mathbf{U},\mathbf{R}) } \right\}. \label{eq:mhratio2}
\end{align}

In the case of a death move, the acceptance probability is calculated as the inverse of \eqref{eq:mhratio2}.

The moves described and constructed in this section give rise to the fully-specified updating algorithm for our proposed single-tree non-ZI count model using the TC sampler, detailed in Algorithm \ref{algo:fullmodelupdatealgo}.

\begin{center}
\begin{algorithm}[H]
 \KwData{Realized observations $(Y_1,\mathbf{X}_1),\dots,(Y_n,\mathbf{X}_n)$ }
 \KwResult{Approximate posterior samples drawn from $\pi(\bm{\lambda},\bm{k},\mathcal{T}, \mathbf{U}, \mathbf{R} \giv (Y_1,\mathbf{X}_1),\dots,(Y_n,\mathbf{X}_n)$)}
 \For{$N_{mcmc}$ iterations}{
	Propose $(\mathcal{T}',\tilde{\bm{\lambda}},\tilde{\bm{k}},\tilde{\mathbf{U}},\tilde{\mathbf{R}}) \giv \cdot \sim q(\mathcal{T}',\tilde{\bm{\lambda}},\tilde{\bm{k}},\tilde{\mathbf{U}},\tilde{\mathbf{R}} \giv \mathcal{T},\bm{\lambda},\bm{k},\mathbf{U},\mathbf{R})$ and accept/reject via a dimension-changing MH step \\
    Draw $(\bm{\lambda}',\mathbf{U}') \giv \tilde{\bm{\lambda}},\tilde{\mathbf{U}},\cdot$ and $(\bm{k}',\mathbf{R}') \giv \tilde{\bm{k}},\tilde{\mathbf{R}},\cdot$ via TC sampler steps
}
 \caption{Posterior sampling algorithm for the proposed single-tree non-ZI count model with taxicab sampler}
 \label{algo:fullmodelupdatealgo}
\end{algorithm}
\end{center}
We followed the ``restart" strategy described in CGM98, running 20 individual chains for 3000 iterations each and restarting each new chain from a single-node tree. 500 burn-in iterations were used for each run and discarded prior to analysis. We utilized $\zeta=50$ cuts to discretize each covariate dimension. Hyperparameter settings for this set of comparison simulations are detailed in Table \ref{table:runtimehp}. 
\begin{table}[t]
\centering
\begin{tabular}{@{} ccc @{}}
  \toprule
    \textbf{Method} & \textbf{Parameters} & \textbf{Values Considered} \\
  \midrule
    \multirow{4}{*}{Na{\"i}ve MH} & $k$ prior: $(\kappa, \beta_k, t_k)$ combinations & (4,1,0.025)  \\
     & Tree depth prior: $(\alpha, \beta)$ combinations & (0.95,4) \\
 & MH proposal radii: $(\lambda,k,c)$ combinations & (4,2,25), (6,2,25) \\
 & Tent pmf tail mass parameter: $t$ & 0.025 \\
 \midrule
    \multirow{4}{*}{Taxicab} & $k$ prior: $(\kappa, \beta_k,t_k)$ combinations & (4,1,0.025) \\
 & Tree depth prior: $(\alpha, \beta)$ combinations & (0.95,4) \\
 & $\mathcal{N}(\cdot)$ radii: $(m_{\lambda},m_k)$ combinations & (2,1), (3,1), (4,2), (5,2) \\
 & MH proposal radius: $c$ & 25 \\
 & Tent pmf tail mass parameter: $t$ & 0.025 \\
  \bottomrule
\end{tabular}
\caption{Hyperparameter settings for runtime comparison.}
\label{table:runtimehp}
\end{table}
The choice of MH proposal radius for cutpoints $c$ was selected so that perturbation proposal corresponding to an existing cut $c=24$ or $25$ could conceivably propose any other available cutpoint value in the corresponding covariate dimension. The choices of ball and MH proposal radii for each $(\lambda_b,k_b)$ pair were intended to highlight any potential differences or variability in fit and computation time. For simplicity we fixed $t=0.025$ and $t_k = 0.025$ for all runs. Assessment of fit was based on mean absolute error (MAE) and $L_2$ norm, both averaged over the 20 runs at each combination of hyperparameter settings. 

We took 1000 posterior samples to compute both the $L_2$ norm and MAE quantities, along with their standard deviation (SD) and standard error (SE) respectively. Total runtime was also recorded at each combination of model settings, measuring the length of time elapsed to execute the model-fitting algorithm for all 20 runs. The results of this comparison are presented in Tables \ref{table:tentpmfexp1results1} and \ref{table:tentpmfexp1results2}, corresponding to the respective outcomes for the na{\"i}ve MH and TC sampler approaches.

\begin{table}[t]
\centering
\begin{tabular}{@{} cccc @{}}
  \toprule
    \textbf{Method} & \textbf{n} & \textbf{$(\lambda,k,c)$ radii} & \textbf{Runtime(sec)} \\
  \midrule
    \multirow{6}{*}{Na{\"i}ve MH} & 10 & (4,2,25) & 101.85 \\
    & 10 & (6,2,25) & 101.81 \\ 
    & 100 & (4,2,25) & 546.45 \\
    & 100 & (6,2,25) & 544.52 \\
    & 1000 & (4,2,25) & 3847.76 \\
    & 1000 & (6,2,25) & 4072.09 \\
   \midrule
     \textbf{Method} & \textbf{n} & \textbf{$(m_{\lambda},m_k,c)$ radii} & \textbf{Runtime(sec)} \\
   \midrule
    \multirow{12}{*}{Taxicab} & 10 & (2,1,25) & 5.11 \\
    & 10 & (3,1,25) & 5.52 \\
    & 10 & (4,2,25) & 7.84 \\ 
    & 10 & (5,2,25) & 8.68 \\ 
    & 100 & (2,1,25) & 24.67 \\
    & 100 & (3,1,25) & 28.93 \\
    & 100 & (4,2,25) & 41.03 \\
    & 100 & (5,2,25) & 53.81 \\
    & 1000 & (2,1,25) & 216.39 \\
    & 1000 & (3,1,25) & 242.84 \\
    & 1000 & (4,2,25) & 337.98 \\
    & 1000 & (5,2,25) & 443.24 \\
  \bottomrule
\end{tabular}
\caption{Comparison of runtime results for models fit with na{\"i}ve MH and TC sampler approaches. All reported values are rounded to the nearest hundredths place.}
\label{table:tentpmfexp1results1}
\end{table}
\begin{table}[t]
\centering
\begin{tabular}{@{} ccccc @{}}
  \toprule
    \textbf{Method} & \textbf{n} & \textbf{$(\lambda,k,c)$ radii} & \textbf{MAE(SE)} & \textbf{$L_2$ norm(SD)} \\
  \midrule
    \multirow{6}{*}{Na{\"i}ve MH} & 10 & (4,2,25) & 6.12(0.02) & 189.11(36.79) \\
    & 10 & (6,2,25) & 6.15(0.02) & 189.39(41.17) \\ 
    & 100 & (4,2,25) & 2.73(0.00) & 52.30(1.88) \\
    & 100 & (6,2,25) & 2.73(0.00) & 52.57(1.97) \\
    & 1000 & (4,2,25) & 2.73(0.00) & 54.22(3.10) \\
    & 1000 & (6,2,25) & 2.75(0.01) & 55.84(11.88) \\
  \midrule
   \textbf{Method} & \textbf{n} & \textbf{$(m_{\lambda},m_k,c)$ radii} & \textbf{MAE(SE)} & \textbf{$L_2$ norm(SD)} \\
  \midrule 
    \multirow{12}{*}{Taxicab} & 
    10 & (2,1,25) & 6.23(0.04) & 195.68(23.43) \\
    & 10 & (3,1,25) & 6.17(0.03) & 193.60(21.77) \\
    & 10 & (4,2,25) & 6.05(0.03) & 192.16(27.89) \\ 
    & 10 & (5,2,25) & 6.08(0.02) & 191.33(21.80) \\ 
    & 100 & (2,1,25) & 3.98(0.42) & 84.00(48.17) \\
    & 100 & (3,1,25) & 3.49(0.26) & 76.63(42.65) \\
    & 100 & (4,2,25) & 2.76(0.23) & 52.84(5.12) \\
    & 100 & (5,2,25) & 2.74(0.00) & 52.33(2.81) \\
    & 1000 & (2,1,25) & 2.75(0.01) & 54.77(1.82) \\
    & 1000 & (3,1,25) & 2.75(0.01) & 54.72(2.03) \\
    & 1000 & (4,2,25) & 2.73(0.00) & 54.80(2.47) \\
    & 1000 & (5,2,25) & 2.74(0.01) & 54.28(1.25) \\
  \bottomrule
\end{tabular}
\caption{Comparison of MAE and $L_2$ norm results for models fit with na{\"i}ve MH and TC sampler approaches. All reported values are rounded to the nearest hundredths place.}
\label{table:tentpmfexp1results2}
\end{table}

\subsection{Performance Comparison}
\label{ss:perfcomp}

Performance between the two methods with respect to $L_2$ norm was comparable across sample size, with improved recovery of the true underlying $(\mathcal{T},\mathbf{M})$ for larger $n$. As expected, computation time increased in both methods with sample size and larger choice of $\mathcal{N}(\cdot)$ radii in the TC sampler approach. Depending on sample size, the TC algorithm was anywhere between 8 to 20 times faster than the na{\"i}ve MH algorithm for ``similar" $\mathcal{N}(\cdot)$ and MH proposal radii settings. The runtime improvements were on the larger end of this range for $(m_{\lambda},m_k) \in \{(2,1),(3,1)\}$, though the reported MAE and $L_2$ norm values at these settings were suboptimal compared to the $(m_{\lambda},m_k) = (4,2)$ and $(5,2)$ settings at the $n=100$ sample size. With respect to TC sampler results for the $n=1000$ sample size, the most probable tree configurations were frequently close to the ground truth tree structure, with some runs identifying somewhat larger trees due to the inclusion of extraneous internal node rules that were unable to be pruned away; we note that similar behavior occurred with most probable tree configurations at this sample size setting under the na{\"i}ve MH sampler with comparable frequency, indicating the ``excess" estimated tree structure is an effect of the underlying stochastic-search mechanism used to explore posterior trees in these kinds of models, as opposed to an inherent issue with the TC sampler itself.

Though the calculated MAEs were in-sample, they are included in both summary tables as a simple way to screen any potentially noticeable differences in mean parameter fits both within and between model fits according to the two algorithms. The MAEs for the $(m_{\lambda},m_k)=(2,1)$ and $(m_{\lambda},m_k)=(3,1)$ settings at $n=100$ are somewhat higher than their counterparts under the na{\"i}ve MH sampler, suggesting that chains involving combinations of smaller $(m_{\lambda},m_k)$ values require longer mixing time for intermediate sample sizes due to constraints imposed in the construction of our dimension-changing proposals; however, the MAE in the $(m_{\lambda},m_k)=(4,2)$ and $(5,2)$ settings at $n=100$ is in line with the $n=100$ results in the na{\"i}ve MH sampler, indicating that, in medium sample size cases, the TC sampler is able to identify good tree structures and terminal node parameter values in shorter chains at other reasonable $\mathcal{N}(\cdot)$ radii settings. Otherwise, MAEs did not appear appreciably different between the two approaches in the reported results, showing that the TC sampler performs comparably to the na{\"i}ve MH sampler in a large number of cases at a fraction of the computation time.

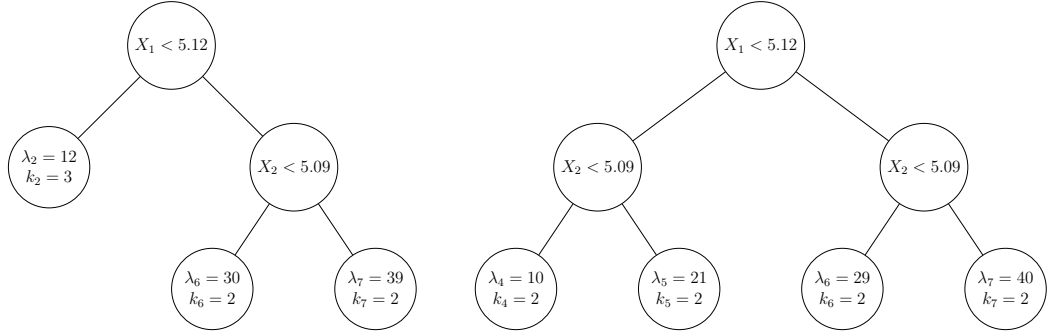
\begin{figure}[t]
\centering
    \begin{minipage}[c]{0.5\linewidth}
    \centering
    \begin{tikzpicture}[scale=0.54, every tree
    node/.style={draw,circle,minimum size=2cm,align=center},
       level distance=3cm,sibling distance=2cm, 
       edge from parent path={(\tikzparentnode) -- (\tikzchildnode)}]
    \Tree [.{$X_{1} < 5.12$} 
    	[.{$\lambda_2 = 12$\\$k_2 = 3$} ]
        [.\node{$X_{2} < 5.09$};
           \edge; [.\node{$\lambda_6 = 30$\\$k_6 = 2$}; ] \edge; [.\node{$\lambda_7 = 39$\\$k_7 = 2$}; ]
        ]
     ]
    \end{tikzpicture}
    \end{minipage}%
    \begin{minipage}[c]{0.5\linewidth}
    \centering
    \begin{tikzpicture}[scale=0.54,every tree
    node/.style={draw,circle,minimum size=2cm,align=center},
       level distance=3cm,sibling distance=2cm, 
       edge from parent path={(\tikzparentnode) -- (\tikzchildnode)}]
    \Tree [.{$X_{1} < 5.12$} 
        [.{$X_2 < 5.09$}
            {$\lambda_4 = 10$\\$k_4 = 2$} {$\lambda_5 = 21$\\$k_5 = 2$}
        ]
        [.{$X_2 < 5.09$} 
            {$\lambda_6 = 29$\\$k_6 = 2$} {$\lambda_7 = 40$\\$k_7 = 2$}    
        ]
     ]
    \end{tikzpicture}
    \end{minipage}
\caption{Left: most-probable tree configuration identified on run \#5  (with 26.13\% within-run posterior probability) using TC sampler algorithm at $n=100$ with $(m_{\lambda},m_k) = (2,1)$ and $L_2$ norm = 137.397; the displayed terminal node parameter values are from a single saved iteration associated with this tree structure. In this configuration the sampler has not accepted a birth proposal with internal rule involving $X_2$ at node $\eta_2$ to recover the optimal tree structure given the generated cutpoints. Right: most-probable tree configuration identified on run \#1 (with 31.17\% within-run posterior probability) using TC sampler algorithm at $n=100$ with $(m_{\lambda},m_k) = (2,1)$ and $L_2$ norm = 50.00.}
\label{fig:mptreefig}
\end{figure}

\section{Discussion}
\label{sec:discussion}

The taxicab sampler presented here builds on the ideas presented in \cite{titsias_2017}, offering a flexible, natural  and useful extension to operations on non-binary discrete state spaces in Bayesian models. We have demonstrated the improved efficiency and inferential capabilities of the TC sampler relative to a MH sampler in a challenging univariate setting involving a complicated multimodal distribution, suggesting that even equipping the TC sampler with relatively small choices of radius parameter $m$ can result in tangible performance improvements over a MH sampler with comparable choice of random walk proposal radius value. 

Here we have also shown the ability of the TC sampler to aid in performing efficient inference in a Bayesian regression tree count model setting with comparable performance to that of a na{\"i}ve MH sampler implementation at a fraction of the computational cost; further gains in speed could be achieved by parallelizing TC sampler computations. Further, while exotic, our proposed single-tree count model offers a number of advantages, including interpretability of model parameters and the ability to readily model under-, equi-, and over-dispersion over different regions of covariate space, showing how the use of discrete parameter spaces in tandem with a non-conjugate, non-exponential-family-based model specification can serve as an interesting alternative compared to a more traditional modeling approach relying on continuous latent state spaces.

The TC sampler may also offer additional benefits in the context of efficiently searching over the posterior tree space in Bayesian regression tree models. Whereas \cite{mohammadi_2020} use a full marginalization strategy to perform advantageous tree updates in these types of models, the ability to marginalize over a subset of tree structures to sidestep traversing low-probability regions of the posterior tree space is desirable in its own right, and the local marginalization approach presented in this paper offers interesting considerations for this tree-mixing problem if a suitable distance can be identified.

\bibliographystyle{plainnat}
\bibliography{bibliography}

\newpage

\begin{appendices}

\section{Taxicab Sampler Proofs}
\label{appdx:taxicabsamplerproofs}

\subsection{Detailed Balance Condition for the Taxicab Sampler}
\label{appdx:detailedbalanceproof}

\cite{liu_1994} provide the general proof of a detailed balance condition for data-augmented Markov chains. We show that this result holds with respect to arbitrary conditional posterior distributions of interest in models containing discrete parameters in our taxicab sampler framework. For simplicity, we show the result holds in the univariate setting. Note that for the results in this and the next subsection to hold, we require $m_{\lambda} \geq 1$ and, as in Section \ref{sec:taxicab-sampler} of the main text, we assume the discrete parameter $\lambda$ and its associated auxiliary variable $U$ are of the same dimension. For compactness we suppress dependencies on $\mathbf{X}$ and work with the expression

\begin{equation*}
    p(\mathbf{Y},\bm{\theta},U) = \sum_{\lambda^* \in \mathcal{N}_{m_{\lambda}}(U)} p(\mathbf{Y},\bm{\theta},\lambda^*).
\end{equation*}

We begin by marginalizing the bivariate kernel $q_{m_{\lambda}}(U^{(t+1)},\lambda^{(t+1)} \giv U^{(t)},\lambda^{(t)})$ over $U$ to obtain a marginal transition kernel in $\lambda$, which we denote by $q_{m_{\lambda},\lambda}$, such that
\begin{align}
\begin{split}
    q_{m_{\lambda},\lambda}(\lambda^{(t+1)} \giv \lambda^{(t)}) &= \sum_{U^{(t+1)} \in \Z} q_{m_{\lambda}}(U^{(t+1)},\lambda^{(t+1)} \giv U^{(t)},\lambda^{(t)}) \\
    &= \sum_{U^{(t+1)} \in \Z} p(U^{(t+1)} \giv \lambda^{(t)})q_{m_{\lambda}}(\lambda^{(t+1)} \giv \mathbf{Y},\bm{\theta},U^{(t+1)}).
\end{split}
\label{appdxeq:lambdamargkernel}
\end{align}
The marginal kernel in \eqref{appdxeq:lambdamargkernel} induces a chain $\{\lambda^{(t)},t=0,1,2,... \}$, and for $t \geq 0$ we next show the chain satisfies the detailed balance condition with respect to a stationary distribution $\pi(\lambda|\mathbf{Y},\bm{\theta})$: 
\begin{align}
\label{appdxeq:detbal}
    &\pi(\lambda^{(t)}\giv\mathbf{Y},\bm{\theta})q_{m_{\lambda},\lambda}(\lambda^{(t+1)}\giv\lambda^{(t)}) \\
    &= \sum_{U^{(t+1)} \in \Z} \pi(\lambda^{(t)}\giv\mathbf{Y},\bm{\theta}) q_{m_{\lambda}}(U^{(t+1)},\lambda^{(t+1)} \giv U^{(t)},\lambda^{(t)}) \nonumber \\
    &= \sum_{U^{(t+1)} \in \Z} \frac{p(\mathbf{Y},\bm{\theta},\lambda^{(t)})}{p(\mathbf{Y},\bm{\theta})} \frac{p(\mathbf{Y},\bm{\theta},\lambda^{(t+1)})\mathbb{I}\{\lambda^{(t+1)} \in \mathcal{N}_{m_{\lambda}}(U^{(t+1)}) \}}{p(\mathbf{Y},\bm{\theta},U^{(t+1)})}  \frac{\mathbb{I}\{U^{(t+1)} \in \mathcal{N}_{m_{\lambda}}(\lambda^{(t)})\}}{Z_{m_{\lambda}}} \nonumber \\
    &= \sum_{\substack{U^{(t+1)} \in \mathcal{N}_{m_{\lambda}}(\lambda^{(t)}) \\ \wedge \ U^{(t+1)} \in \mathcal{N}_{m_{\lambda}}(\lambda^{(t+1)})} } \frac{p(\mathbf{Y},\bm{\theta},\lambda^{(t)})}{p(\mathbf{Y},\bm{\theta})} \frac{p(\mathbf{Y},\bm{\theta},\lambda^{(t+1)})\mathbb{I}\{\lambda^{(t+1)} \in \mathcal{N}_{m_{\lambda}}(U^{(t+1)}) \}}{p(\mathbf{Y},\bm{\theta},U^{(t+1)})}  \frac{\mathbb{I}\{U^{(t+1)} \in \mathcal{N}_{m_{\lambda}}(\lambda^{(t)})\}}{Z_{m_{\lambda}}} \nonumber \\
    &= \sum_{\substack{U^{(t)} \in \mathcal{N}_{m_{\lambda}}(\lambda^{(t)}) \\ \wedge \ U^{(t)} \in \mathcal{N}_{m_{\lambda}}(\lambda^{(t+1)})} } \frac{p(\mathbf{Y},\bm{\theta},\lambda^{(t+1)})}{p(\mathbf{Y},\bm{\theta})} \frac{p(\mathbf{Y},\bm{\theta},\lambda^{(t)})\mathbb{I}\{\lambda^{(t)} \in \mathcal{N}_{m_{\lambda}}(U^{(t)}) \}}{p(\mathbf{Y},\bm{\theta},U^{(t)})}  \frac{\mathbb{I}\{U^{(t)} \in \mathcal{N}_{m_{\lambda}}(\lambda^{(t+1)})\}}{Z_{m_{\lambda}}} \nonumber \\
    &= \pi(\lambda^{(t+1)}\giv\mathbf{Y},\bm{\theta}) \sum_{U^{(t)} \in \Z} p_{m_{\lambda}}(\lambda^{(t)}\giv\mathbf{Y},\bm{\theta},U^{(t)})p(U^{(t)}\giv\lambda^{(t+1)}) \nonumber \\
    &= \pi(\lambda^{(t+1)}\giv\mathbf{Y},\bm{\theta}) \sum_{U^{(t)} \in \Z} q_{m_{\lambda}}(\lambda^{(t)},U^{(t)}\giv\lambda^{(t+1)},U^{(t+1)}) \nonumber \\
    &= \pi(\lambda^{(t+1)}\giv\mathbf{Y},\bm{\theta})q_{m_{\lambda},\lambda}(\lambda^{(t)}\giv\lambda^{(t+1)}). \nonumber
\end{align}
Here the equality between the third and fourth lines of (\ref{appdxeq:detbal}) holds due to the symmetry of $\mathcal{N}_{m_{\lambda}}(\cdot)$, in that for any given $U$ and $\lambda$, $\mathbb{I}\{U \in \mathcal{N}_{m_{\lambda}}(\lambda)\} = \mathbb{I}\{\lambda \in \mathcal{N}_{m_{\lambda}}(U)\}$ for $U \in \Z,\lambda \in \Z$. Application of this symmetry result also allows us to rewrite the summation over $U^{(t+1)} \in \Z$ in the third line as a summation over all $U^{(t+1)} \in \mathcal{N}_{m_{\lambda}}(\lambda^{(t)}) \wedge \ U^{(t+1)} \in \mathcal{N}_{m_{\lambda}}(\lambda^{(t+1)})$ in the fourth line; we obtain equality between the fourth and fifth lines by recognizing the additional requirement that $U^{(t)} \in \mathcal{N}_{m_{\lambda}}(\lambda^{(t)}) \wedge \ U^{(t)} \in \mathcal{N}_{m_{\lambda}}(\lambda^{(t+1)})$, so that the set elements in the summations are the same between the two lines.

Thus (\ref{appdxeq:detbal}) shows $\pi(\lambda \giv \mathbf{Y},\bm{\theta})$ is indeed a stationary distribution with respect to the marginal chain in $\{ \lambda^{(t)} \}_{t=0}^{\infty}$, since the detailed balance condition in turn demonstrates that
\begin{align*}
\begin{split}
    \sum_{\lambda^{(t+1)} \in \Z} \pi(\lambda^{(t+1)} \giv \mathbf{Y},\bm{\theta})q_{m_{\lambda},\lambda}(\lambda^{(t)} \giv \lambda^{(t+1)}) &= \sum_{\lambda^{(t+1)} \in \Z} \pi(\lambda^{(t)}|\mathbf{Y},\bm{\theta})q_{m_{\lambda},\lambda}(\lambda^{(t+1)} \giv \lambda^{(t)}) \\
    &= \pi(\lambda^{(t)} \giv \mathbf{Y},\bm{\theta}),
\end{split}
\end{align*}
i.e. $\pi(\lambda \giv \mathbf{Y},\bm{\theta})$ satisfies the global balance condition with respect to $q_{m_{\lambda},\lambda}$. The stationarity result implies that, if at some time $t >0$ we have $\lambda^{(t)} \sim \pi(\lambda\giv\mathbf{Y},\bm{\theta})$ in our marginal chain, then all subsequent states $\lambda^{(t+1)},\lambda^{(t+2)},...$ must be distributed according to $\pi(\lambda \giv \mathbf{Y},\bm{\theta})$ also.

\subsection{Irreducibility and Aperiodicity of the Marginal Chain $\{ \lambda^{(t)}\}_{t=0}^{\infty}$}
\label{appdx:stationarydistnproof}

To prove $\pi(\lambda \giv \mathbf{Y},\bm{\theta})$ is indeed the unique and limiting stationary distribution with respect to this marginal chain, we also need to show the chain is irreducible and aperiodic. Here we impose an additional assumption on the state space $\Lambda$, namely that $\Lambda = \{\lambda: p(\lambda \giv \mathbf{Y},\bm{\theta}) > 0, \lambda \in \Z \}$. We note that this is a reasonable assumption for many models in which the prior distribution on $\lambda$ places positive probability on all of $\Z$; the following result holds still in the setting where the prior distribution on $\lambda$ places positive probability on countable/finite subsets of $\Z$. From here, it is straightforward to show that irreducibility holds for the marginal chain in $\lambda$, since for any one-step transition from $\lambda^{(t)} = j$ to $\lambda^{(t+1)} = j+l,\ j \in \Z ,\ l \in \{-1,1 \}$, with ball radius $m_{\lambda}$, we observe that
\begin{align}
    q_{m_{\lambda},\lambda}(\lambda^{(t+1)} &= j+l \giv \lambda^{(t)} = j) \label{appdxeq:irredfirst} \\
    &= \sum_{U^{(t+1)} \in \Z} p(U^{(t+1)}\giv \lambda^{(t)}) q_{m_{\lambda}}(\lambda^{(t+1)}\giv \mathbf{Y},\bm{\theta},U^{(t+1)}) \\
    &= \sum_{U^{(t+1)} \in \Z}  \frac{p(\mathbf{Y},j+l,\bm{\theta})\mathbb{I}\{j+l \in \mathcal{N}_{m_{\lambda}}(U^{(t+1)}) \}}{p(\mathbf{Y},\bm{\theta},U^{(t+1)})}  \frac{\mathbb{I}\{U^{(t+1)} \in \mathcal{N}_{m_{\lambda}}(j)\}}{Z_{m_{\lambda}}} \label{appdxeq:irredthird} \\
    &= \sum_{U^{(t+1)} \in \Z}  \frac{p(\mathbf{Y},j+l,\bm{\theta})\mathbb{I}\{U^{(t+1)} \in \mathcal{N}_{m_{\lambda}}(j+l) \}}{p(\mathbf{Y},\bm{\theta},U^{(t+1)})}  \frac{\mathbb{I}\{U^{(t+1)} \in \mathcal{N}_{m_{\lambda}}(j)\}}{Z_{m_{\lambda}}} \label{appdxeq:irredfourth} \\
    &= \sum_{\substack{U^{(t+1)} \in \mathcal{N}_{m_{\lambda}}(j) \\ \wedge \ U^{(t)} \in \mathcal{N}_{m_{\lambda}}(j+l)} } p(U^{(t+1)}\giv \lambda^{(t)})q_{m_{\lambda}}(\lambda^{(t+1)}\giv \mathbf{Y},\bm{\theta},U^{(t+1)}) \label{appdxeq:irredfifth} \\
    &> 0, \label{appdxeq:irredlast}
\end{align}
since in (\ref{appdxeq:irredthird}) and (\ref{appdxeq:irredfourth})
\begin{equation*}
    p(\mathbf{Y},\bm{\theta},U^{(t+1)}) = \sum_{\lambda^{*} \in \mathcal{N}_{m_{\lambda}}(U^{(t+1)})} p(\mathbf{Y},\bm{\theta},\lambda^{*})
\end{equation*}
is always well-defined and greater than zero due to our assumption on the state space $\Lambda$. We once again obtain (\ref{appdxeq:irredfourth}) through the above-described symmetry property, giving $\mathbb{I}\{j+l \in \mathcal{N}_1(U^{(t+1)})\} = \mathbb{I}\{U^{(t+1)} \in \mathcal{N}_1(j+l)\}$. 

This result shows that all adjacent states $j$ and $j+l$ in $\Lambda$ communicate, showing that the chain is irreducible. 

To see that the chain is aperiodic, we simply let $l=0$ in (\ref{appdxeq:irredfirst})-(\ref{appdxeq:irredlast}), which shows the one-step transition probability $q_{m_{\lambda},\lambda}(\lambda^{(t+1)}=j \giv \lambda^{(t)}=j) > 0$ and so $\text{gcd} \{n \geq 1:q_{m_{\lambda},\lambda}^n(\lambda^{(t+n)}=j \giv \lambda^{(t)}=j) > 0\} = 1$ also.

The detailed balance condition combined with the above irreducibility/aperiodicity results show that the marginal chain is positive recurrent, and that $\pi(\lambda \giv \mathbf{Y},\bm{\theta})$ is indeed both the unique and limiting stationary distribution with respect to the marginal chain $\{\lambda^{(t)},t=0,1,2,... \}$.

\section{Comparison of TC and MH Samplers in a Simulated Example}

In this section we demonstrate the TC sampler's ability to draw approximately from target distributions of interest in settings where a MH sampler fails. 

\subsection{An Example with Infinite Support}
We consider a complicated discrete distribution $p(\lambda)$ for some parameter $\lambda \in \Z_{\geq 0}$, where $p(\lambda)$ places mass primarily on odd-valued elements of $\Z_{\geq 0}$. Such an example is intended to simulate a more challenging setting in which the underlying state space is highly multimodal with infinite but countable support. Here we modify a Poisson distribution to write $p(\lambda)$ as
\begin{align}
    p(\lambda) \propto
    \begin{cases}
    w \frac{10^{\lambda}e^{-10}}{\lambda!}, & \text{ if } \lambda = 2k, k \in \Z_{\geq 0}, \\
    (1-w) \frac{10^{\lambda}e^{-10}}{\lambda!}, & \text{ if } \lambda = 2k+1, k \in \Z_{\geq 0}.
    \end{cases}
    \label{eq:plambda_inf_support}
\end{align}

For our purposes we take $w=0.0005,$ resulting in 0.1\% of the total mass under $p(\lambda)$ being assigned to even non-negative integers and the remaining mass assigned to odd non-negative integers.

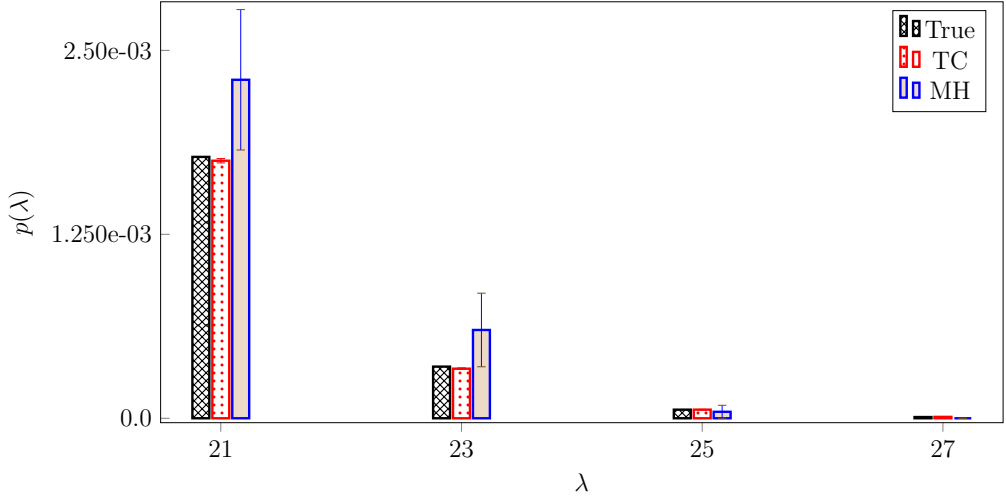
\begin{figure}
\centering
\begin{tikzpicture}[scale = 0.8]
 
\begin{axis} [ybar=.05cm,
    bar width = 8pt,
    xlabel=$\lambda$,
	ylabel=$p(\lambda)$,
    ymin=0.0,
    ymax=2.8e-03,
    ytick={0.0,1.250e-03,2.50e-03},
	yticklabels={0.0,1.250e-03,2.50e-03},
	yticklabel style={
        /pgf/number format/fixed,
        /pgf/number format/precision=2
    },
    scaled y ticks=false,
    yticklabel style={
        /pgf/number format/fixed,
        /pgf/number format/precision=2
    },
    enlarge y limits = {abs = .00003},
    enlarge x limits = {abs = .5},
    xtick=data,
    width=10cm,height=7cm,
    x=2cm,
    scale only axis,
    xtick pos =left,
    ytick pos =left
]
 
\addplot[draw=black,pattern=crosshatch,pattern color=black,line width=0.4mm] coordinates {
	(21, 1.776e-03)
	(23, 3.51e-04)
	(25, 5.85e-05)
	(27, 8.33e-06)
};
 
\addplot+[color=red,pattern=dots,pattern color=red,line width=0.4mm,error bars/.cd,
y dir=both,y explicit] coordinates {
    (21, 1.75e-03) +- (0,1.47e-05)
	(23, 3.37e-04) +- (0,5.43e-06)
	(25, 5.86e-05) +- (0,1.96e-06)
	(27, 9.51e-06) +- (0,7.22e-07)
};

\addplot+[draw=blue,line width=0.4mm,error bars/.cd,
y dir=both,y explicit] coordinates {
    (21, 0.0023) +- (0,0.000477)
	(23, 6.00e-04) +- (0,0.00025)
	(25, 4.41e-05) +- (0,4.412e-05)
	(27, 0) +- (0,0)
};

\legend{True,TC,MH}
\end{axis}
\end{tikzpicture}
\caption{Mean empirical probabilities (with standard errors) of visiting select odd-valued states under the TC (red dotted bars) and MH samplers (blue bars) after 1,000,000 iterations compared with true probabilities under $p(\lambda)$ (black crosshatched bars). Displayed values were computed based on 100 samples.}
\label{fig:mh_vs_ts_right_tail}
\end{figure}

\begin{figure}
\centering
\begin{tikzpicture}[scale=0.8]
 
\begin{axis} [ybar=.05cm,
    bar width = 8pt,
    xlabel=$\lambda$,
	ylabel=$p(\lambda)$,
    ymin=0.0,
    ymax=2.60e-06,
    ytick={0.0,1.00e-06,2.00e-06},
	yticklabels={0.0,1.00e-06,2.00e-06},
	yticklabel style={
        /pgf/number format/fixed,
        /pgf/number format/precision=2
    },
    scaled y ticks=false,
    yticklabel style={
        /pgf/number format/fixed,
        /pgf/number format/precision=2
    },
    enlarge y limits = {abs = .00000003},
    enlarge x limits = {abs = .5},
    xtick=data,
    width=10cm,height=7cm,
    x=2cm,
    scale only axis,
    xtick pos =left,
    ytick pos =left
]
 
\addplot[draw=black,pattern=crosshatch,pattern color=black,line width=0.4mm] coordinates {
	(20, 1.87e-06)
	(22, 4.04e-07)
	(24, 7.32e-08)
	(26, 1.13e-08)
};
 
\addplot+[color=red,pattern=dots,pattern color=red,line width=0.4mm,error bars/.cd,
y dir=both,y explicit] coordinates {
	(20, 2.01e-06) +- (0,1.33e-07)
	(22, 3.70e-07) +- (0,6.14e-08)
	(24, 7.00e-08) +- (0,2.56e-08)
	(26, 2.00e-08) +- (0,1.407e-08)
};

\addplot+[draw=blue,line width=0.4mm,error bars/.cd,
y dir=both,y explicit] coordinates {
	(20, 2.19e-06) +- (0,3.71e-07)
	(22, 5.6e-07) +- (0,1.79e-07)
	(24, 1.2e-07) +- (0,8.32e-08)
	(26, 0) +- (0,0)
};

\legend{True,TC,MH}
\end{axis}
\end{tikzpicture}
\caption{Mean empirical probabilities (with standard errors) of visiting select even-valued states under the TC (red dotted bars) and MH samplers (blue bars) after 1,000,000 iterations compared with true probabilities under $p(\lambda)$ (black crosshatched bars). Displayed values were computed based on 100 samples.}
\label{fig:mh_vs_ts_right_tail_2}
\end{figure}
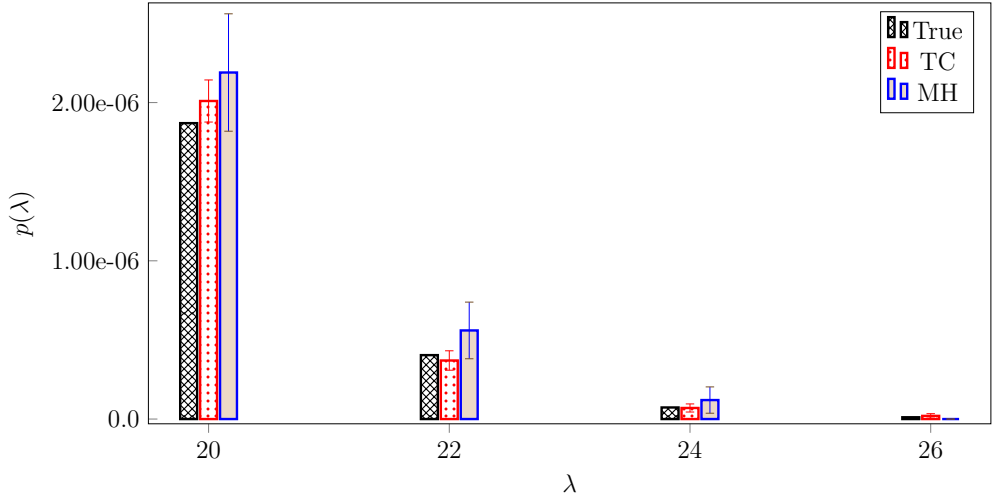

\begin{figure}
\centering
\begin{tikzpicture}
\begin{axis}[
	xlabel=Iterations,
	ylabel=Total Variation Distance,
	xticklabels={},
	yticklabels={},
    extra x ticks={1,2,...,6},
    extra y ticks={0.0,0.2,0.4,0.6,0.8,1.0},
    extra x tick labels={100,1000,10000,100000,1000000},
    extra y tick labels={0.0,0.2,0.4,0.6,0.8,1.0},
	width=10cm,height=7cm,
    legend style={at={(1,0.5)},anchor=east}]
\addplot[color=red,mark=x] coordinates {
	(1, 0.153) 
	(2, 0.050) 
	(3, 0.017) 
	(4, 0.005) 
	(5, 0.002) 
};

\addplot[color=blue,mark=*] coordinates {
	(1, 0.903) 
	(2, 0.833) 
	(3, 0.506) 
	(4, 0.190) 
	(5, 0.061) 
};

\legend{TC,MH}
\end{axis}
\end{tikzpicture}
\caption{Mean total variation distance (based on 100 samples) at selected iterations between estimated and target distributions under the TC and MH samplers (equipped with $m=1$ and proposal radius 1 respectively) for the infinite support example.}
\label{fig:tc_vs_mh_mean_TV_m1}
\end{figure}
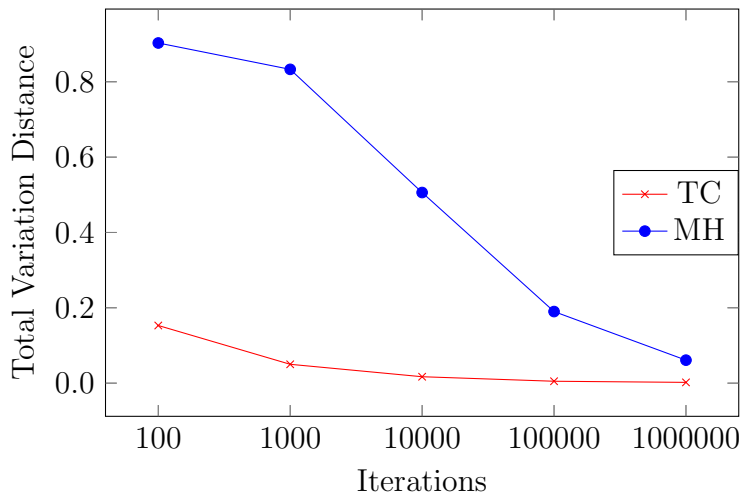

We construct two samplers to explore $p(\lambda)$: a MH sampler using a random walk proposal with radius 1, and a TC sampler with radius $m=1$. This choice of $m$ is not essential in this comparison, as we obtained strikingly similar results with larger values of $m$ as well. We ran 100 chains of length 1,000,000 under each sampler, with each TC-MH pair of chains initialized to a random starting state between 0 and 20. We subsequently reported the mean total variation (TV) and Hellinger (HE) distances for the resulting estimated pmfs compared to the true underlying distribution \eqref{eq:plambda_inf_support}. In addition, we compute the mean TV and HE distances with respect to the aforementioned pmfs conditional on $\lambda$ being either odd or even in order to assess how well the samplers explore the two subspaces. Note that in this experiment, we observe both even and odd states with vanishingly small probabilities as $\lambda \rightarrow \infty$ under \eqref{eq:plambda_inf_support}. As a result, while we do not expect either sampler to fully and adequately explore the corresponding state space within the designated number of iterations, we should expect that both samplers be able to efficiently sample the target distribution for small values of $\lambda$. Instead, we observe a tangible difference between standard MH sampling compared to that of the TC sampler. The results of the experiment are reported in Tables \ref{appdx:tab_mh_ts_dist_2} and \ref{appdx:tab_mh_ts_dist_3}, with additional relevant visualizations provided in Figures \ref{fig:mh_vs_ts_right_tail}, \ref{fig:mh_vs_ts_right_tail_2}, and \ref{fig:tc_vs_mh_mean_TV_m1}.

\begin{table}[!htbp]
\small
\centering
\begin{tabular}{ |c|c|c|c|c| } 
\hline
Sampler & Iteration & Mean TV(SE) & Mean TV.even(SE) & Mean TV.odd(SE) \\
\hline
\multirow{ 5}{*}{TC} & 100 & 0.153(0.007) & 0.006(0.0005) & 0.153(0.007) \\
& 1000 & 0.050(0.002) & 0.0008(0.00005) & 0.050(0.002) \\
& 10000 & 0.017(0.0007) & 0.0002(0.000008) & 0.017(0.0007) \\
& 100000 & 0.005(0.0002) & 0.00005(0.000002) & 0.005(0.0002) \\
& 1000000 & 0.002(0.00005) & 0.00002(0.0000006) & 0.002(0.00005) \\
\hline
\multirow{ 5}{*}{MH} & 100 & 0.903(0.008) & 0.006(0.0005) & 0.903(0.008) \\
& 1000 & 0.833(0.02) & 0.001(0.00006) & 0.833(0.02) \\
& 10000 & 0.506(0.02) & 0.0003(0.00001) & 0.506(0.02) \\
& 100000 & 0.190(0.008) & 0.00009(0.000004) & 0.190(0.008) \\
& 1000000 & 0.061(0.002) & 0.00003(0.000001) & 0.061(0.002) \\
\hline
\end{tabular}
\caption{Mean total variation (TV) distances between the reported $\hat{p}(\lambda)$ distributions obtained under the TC and MH samplers compared to the true underlying distribution \eqref{eq:plambda_inf_support}, with standard errors reported in parentheses. ``TV.even" and ``TV.odd" correspond to TV distances computed with respect to even- and odd-valued states respectively. Reported mean values are rounded to three significant digits except in cases where additional precision is required to obtain a leading non-zero term; reported standard errors are rounded to the first non-zero term.}
\label{appdx:tab_mh_ts_dist_2}
\end{table}

\begin{table}[!htbp]
\small
\centering
\begin{tabular}{ |c|c|c|c|c| } 
\hline
Sampler & Iteration & Mean HE(SE) & Mean HE.even(SE) & Mean HE.odd(SE)  \\
\hline
\multirow{ 5}{*}{TC} & 100 & 0.301(0.009) & 0.048(0.003) & 0.296(0.009) \\
& 1000 & 0.109(0.003) & 0.024(0.0006) & 0.106(0.003) \\
& 10000 &  0.038(0.001) & 0.012(0.0002)  & 0.036(0.001) \\
& 100000 &  0.012(0.0003) & 0.004(0.00008) & 0.012(0.0003) \\
& 1000000 &  0.004(0.0001) & 0.001(0.00003) & 0.004(0.0001) \\
\hline
\multirow{ 5}{*}{MH} & 100 &  0.859(0.009) & 0.049(0.003) & 0.858(0.009) \\
& 1000 &  0.837(0.01) & 0.024(0.0007) & 0.836(0.01) \\
& 10000 &  0.704(0.01) & 0.017(0.0004) & 0.704(0.01) \\
& 100000 &  0.342(0.01) & 0.008(0.0002) & 0.342(0.01) \\
& 1000000 &  0.121(0.003) & 0.003(0.00007) & 0.121(0.003) \\
\hline
\end{tabular}
\caption{Mean Hellinger (HE) distances between the reported $\hat{p}(\lambda)$ distributions obtained under the TC and MH samplers compared to the true underlying distribution \eqref{eq:plambda_inf_support}, with standard errors reported in parentheses. ``HE.even" and ``HE.odd" correspond to HE distances computed with respect to even- and odd-valued states respectively. Reported mean values are rounded to three significant digits except in cases where additional precision is required to obtain a leading non-zero term; reported standard errors are rounded to the first non-zero term.}
\label{appdx:tab_mh_ts_dist_3}
\end{table}

While on average the MH chains in this case appears to steadily converge to $p(\lambda)$ under both the total variation and Hellinger distances in Tables \ref{appdx:tab_mh_ts_dist_2} and \ref{appdx:tab_mh_ts_dist_3}, we note not only that the TC chains on average converge much more quickly under both distances, but also that the still-large mean Hellinger distance associated with the MH chains after 1,000,000 iterations suggests that the corresponding sampler has failed to explore relatively higher-valued states with appropriate frequency. The largest-valued state visited among the 100 MH chains was $\lambda=25$, compared to $\lambda=31$ for the TC chains. Furthermore, the experimental results show that the MH sampler encounters difficulty exploring odd-valued states with appropriate visitation frequencies due to the multimodal behavior of $p(\lambda)$. In contrast, the TC sampler explores larger odd-valued states quite well, with mean frequencies far closer to the true probabilities under \eqref{eq:plambda_inf_support} as observed in Figure \ref{fig:mh_vs_ts_right_tail_2}. 

In general, after 1,000,000 iterations the TC chains have explored both even- and odd-valued states with reasonable frequencies relative to the true distribution $p(\lambda)$, even for the larger even-valued states visited as seen in Figure \ref{fig:mh_vs_ts_right_tail}. The average difference in largest state visited between the two samplers was 7.79, and this behavior in tandem with the other results indicates the TC sampler explores the underlying state space more efficiently than the MH sampler as well. These results support the notion that the TC sampler serves as a competitive alternative to a MH sampler even in relatively challenging scenarios with small choice of $m$.

\section{Dimension-Changing Moves in the Augmented Single-Tree Model}
\label{appdx:rjmcmc}

\subsection{Metropolis-Hastings Acceptance Probability}

We provide here the details concerning the simplified acceptance probability calculation in \eqref{eq:dimchangemhratiosimplified}. Suppressing all dependencies on $\mathbf{X}$, we have

\begin{align*}
    A &= \frac{\pi(\bm{\theta}',\bm{\lambda}',\mathbf{U}',B' \giv \mathbf{Y})}{\pi(\bm{\theta},\bm{\lambda},\mathbf{U},B \giv \mathbf{Y})} \times \frac{q(B,\bm{\theta} \giv B',\bm{\theta}')}{q(B',\bm{\theta}' \giv B,\bm{\theta})} \times \frac{q_{m_{\lambda}}(\bm{\lambda} \giv \bfY, \bm{\theta}, \mathbf{U}, B)}{ p(\mathbf{a})q_{m_{\lambda}}(\bm{\lambda}' \giv \bfY, \bm{\theta}', \mathbf{U}', B') } \\
    &=
    \begin{aligned}[t]
    &\frac{ p(\bfY \giv \bm{\theta}', \bm{\lambda}',B') \pi(\bm{\theta}', \bm{\lambda}',B') p(\mathbf{U}' \giv \bm{\lambda}')}{p(\bfY \giv \bm{\theta}, \bm{\lambda},B) \pi(\bm{\theta}, \bm{\lambda},B) p(\mathbf{U} \giv \bm{\lambda})} \times \frac{q(B,\bm{\theta} \giv B',\bm{\theta}')}{q(B',\bm{\theta}' \giv B,\bm{\theta})} \\
    &\times \frac{p(\bfY \giv \bm{\theta}, \bm{\lambda}, B) \pi(\bm{\theta}, \bm{\lambda},B) \mathbb{I}\{ \bm{\lambda} \in \mathcal{N}_{m_{\lambda}}(\mathbf{U}) \} }{p(\mathbf{a})p(\bfY \giv \bm{\theta}', \bm{\lambda}',B') \pi(\bm{\theta}', \bm{\lambda}',B') \mathbb{I}\{ \bm{\lambda}' \in \mathcal{N}_{m_{\lambda}}(\mathbf{U}') \} } \\
    &\times \frac{\displaystyle \sum_{\bm{\lambda}^* \in \mathcal{N}_{m_{\lambda}}(\mathbf{U}')} p(\bfY,\bm{\theta}',\bm{\lambda}^*,B')}{\displaystyle \sum_{\tilde{\bm{\lambda}} \in \mathcal{N}_{m_{\lambda}}(\mathbf{U})} p(\bfY,\bm{\theta},\tilde{\bm{\lambda}},B)}
    \end{aligned} \\
    &= \frac{\left[\displaystyle \sum_{\bm{\lambda}^* \in \mathcal{N}_{m_{\lambda}}(\mathbf{U}')} p(\mathbf{Y}|\bm{\theta}',\bm{\lambda}^*,B')\pi(\bm{\theta}',\bm{\lambda}^*,B')\right]p(\mathbf{U}'|\bm{\lambda}')q(B,\bm{\theta}|B',\bm{\theta}') }{\left[\displaystyle \sum_{\tilde{\bm{\lambda}} \in \mathcal{N}_{m_{\lambda}}(\mathbf{U})} p(\mathbf{Y}|\bm{\theta},\tilde{\bm{\lambda}},B)\pi(\bm{\theta},\tilde{\bm{\lambda}},B)\right]p(\mathbf{U}|\bm{\lambda})p(\mathbf{a})q(B',\bm{\theta}'|B,\bm{\theta})}.
\end{align*}

\subsection{Death Move Proposals in the Single-Tree Model}

Here using the same notation as in Subsection \ref{ss:runtimesetup}, in the death move forward proposal we are concerned with the following conditional transitions:
\begin{align}
    (\lambda_{b(l)},\lambda_{b(r)}) &\rightarrow U_b \label{appdx:lambdadeath1} \\
    U_b &\rightarrow \lambda_b \label{appdx:lambdadeath2} \\
(k_{b(l)},k_{b(l)}) &\rightarrow R_b \label{appdx:kdeath1} \\
R_b &\rightarrow k_b \label{appdx:kdeath2}
\end{align}
given $\mathcal{T}'$ and $\mathcal{T}$. 

We obtain the transitions \eqref{appdx:lambdadeath1} and \eqref{appdx:kdeath1} deterministically via $\delta^{-1}$, such that $U_b = \delta^{-1}_1(\lambda_{b(l)},\lambda_{b(r)})$ and $R_b = \delta^{-1}_1(k_{b(l)},k_{b(r)})$. \eqref{appdx:lambdadeath2} and \eqref{appdx:kdeath2} are then proposed via \eqref{eq:balllambdaq} and \eqref{eq:ballkq}.

The death move reverse proposal requires accounting for the following (conditional) transitions:

\begin{align}
    \lambda_b &\rightarrow (U_{b(l)},U_{b(r)}) \label{appdx:lambdadeathrev1} \\
    (U_{b(l)},U_{b(r)}) &\rightarrow (\lambda_{b(l)},\lambda_{b(r)}) \label{appdx:lambdadeathrev2} \\
    k_b &\rightarrow (R_{b(l)},R_{b(l)}) \label{appdx:kdeathrev1} \\
    (R_{b(l)},R_{b(l)}) &\rightarrow (k_{b(l)},k_{b(r)}) \label{appdx:kdeathrev2}
\end{align}

We account for the dimension-matching scalars $a_{\lambda}$ and $a_k$ in this reverse direction by taking $a_{\lambda} = \delta^{-1}_2(\lambda_{b(l)},\lambda_{b(r)})$ and $a_k = \delta^{-1}_2(k_{b(l)},k_{b(r)})$. Finally, \eqref{appdx:lambdadeathrev1}-\eqref{appdx:kdeathrev2} are handled according to the birth move proposal procedure described in Subsection \ref{ss:runtimesetup}.

\section{Hyperparameter Selection}
\label{appdx:hp-calibration}

Here we detail considerations for hyperparameter selection in our non-ZI count model, both for the location and scale prior distributions.

\subsection{Location Prior Distribution}

As described in the main manuscript, for each $\lambda_b \simind \text{DU}\{d_1,d_2 \}$ we have specified $d_1 = \min(\mathbf{Y})$ and $d_2 = \min(\mathbf{Y})$ as reasonable but vague hyperparameter values, with the primary motivation being the assignment of prior probability to a wide range of plausible location parameter values based on the observed data.

\subsection{Tent PMF Tail Mass Parameter and Scale Prior Distribution}

We combine discussion surrounding initial selection of the tent pmf tail mass parameter $t$ and the scale prior distribution location hyperparameter $\kappa$, as the tail definition for our tent pmf leads to a relationship between the likelihood function parameters $t$ and $k$ that makes their joint estimation more tractable compared to estimation of the two parameters separately.

Here we take the approach advocated by \cite{oakley_2002} for eliciting hyperparameters, wherein the analyst selects a prior distribution, generates data from resulting prior predictive distribution based on fixed candidate hyperparameter values, and calibrates the hyperparameters accordingly in tandem with an expert. In the absence of a subject-matter expert or in settings where the analyst seeks candidate hyperparameter values as starting points for the calibration process, we may derive candidate $(t,\kappa)$ combinations assuming the choice of likelihood function and prior distributions specified in the main manuscript.

Explicitly, we assume

\begin{align*}
    Y_i \giv \lambda,k,t,\mathcal{T} &\simind P_t(\lambda_i,\floor{e^{k_i}}), \\
    \lambda_i \giv \mathcal{T} &\simiid \text{DU}\{d_1,d_2 \}, \\
    k_i \giv \mathcal{T},\lambda_i,t_k &\simind P_{t_k}(\floor{\kappa/2^{d_i}},\floor{\widetilde{\log}(\lambda_i)/(1+d_i)^{\beta_k}}), \\
    \mathcal{T} &\sim \pi(\mathcal{T}|\alpha,\beta,\zeta_v),
\end{align*}

so that at the grand mean level of the model, under this choice of scale prior distribution we on average expect the relationship

\begin{align}
    &\hat{y}_{1-t} - m \approx \floor{e^{\floor{\kappa/2^d}}} \nonumber \\
    \rightarrow\ &\hat{y}_{1-t} - m \leq e^{\floor{\kappa/2^d}} < \hat{y}_{1-t} - m + 1,
    \label{appdx:t_k_rel}
\end{align}

where $\hat{y}_{1-t}$ denotes the sample $(1-t)$th quantile and $m$ denotes the sample median. Note that if we believe the specified model is in fact the true data-generating process, we need also estimate $d$, the terminal node depth at which observations are being generated. A point estimate $\hat{d}$ for $d$ may be obtained by drawing a large number of trees from the tree prior distribution for fixed $(\alpha,\beta)$ and obtaining a corresponding summary statistic for the drawn tree terminal node depths, e.g. a mean.

Thus, plugging in $\hat{d}$ for $d$ in \eqref{appdx:t_k_rel} provides a means for estimating $\kappa$ given $t$ by selecting a value $\hat{\kappa}$ uniformly from the set of $\kappa$ values satisfying \eqref{appdx:t_k_rel}; however, the inequality \eqref{appdx:t_k_rel} does not always guarantee a solution in $\kappa$, and in such cases we propose choosing the values satisfying $\argmax_{\kappa} \exp(\floor{\kappa / 2^{\hat{d}}}) < \hat{y}_{1-t} - m$ and $\argmin_{\kappa} \hat{y}_{1-t} - m + 1 \leq \exp(\floor{\kappa / 2^{\hat{d}}})$ and selecting an estimate $\hat{\kappa}$ uniformly from the resulting candidate values. In the case where $\hat{y}_{1-t} - m \leq 1$, we further propose drawing $\hat{\kappa} \sim \text{Bern}(0.5)$ due to the presence of the floor function in \eqref{appdx:t_k_rel}. The above suggests a procedure for selecting $\kappa$ given $t$ as laid out in Algorithm \ref{algo:pseudokappa}.
\begin{algorithm}[t]
 \KwData{Simulated data $(y_1,...,y_n)$, known tail mass hyperparameter $t$, estimate $m$ of median $\lambda$, estimate of terminal node depth $\hat{d}$, estimate of upper (1-t)th quantile $\hat{y}_{1-t}$}
 \KwResult{An estimate $\hat{\kappa}$ of $\kappa$}
 Set $\hat{\kappa} \leftarrow 0$\;
 Calculate $\hat{y}_{1-t}-m,\hat{y}_{1-t}-m+1$ \;
 \eIf{ $(\hat{y}_{1-t}-m) \leq 1$}{
    Draw $\hat{\kappa} \sim \text{Bern}(0.5)$\;
 }{
    \eIf{ $\exists x \in \mathbb{Z}_{\geq 0} : \hat{y}_{1-t}-m \leq \exp{(\floor{x/2^{\hat{d}}})} < \hat{y}_{1-t}-m+1$  }{
        $X \leftarrow \{x \in \mathbb{Z}_{\geq 0}: \hat{y}_{1-t}-m \leq \exp{(\floor{x/2^{\hat{d}}})} < \hat{y}_{1-t}-m+1 \} $\;
        $\hat{\kappa} \leftarrow \text{sample}(X,size=1,replace=\text{FALSE})$\;
    }{
        $X \leftarrow \operatorname*{argmax}_{x \in \mathbb{Z}_{\geq 0}} \exp{(\floor{x/2^{\hat{d}}})} < \hat{y}_{1-t}-m+1$\;
        $Y \leftarrow \operatorname*{argmin}_{x \in \mathbb{Z}_{\geq 0}} \hat{y}_{1-t}-m+1 \leq \exp{(\floor{x/2^{\hat{d}}})}$\;
        $\hat{\kappa} \leftarrow \text{sample}(X \cup Y,size=1,replace=\text{FALSE})$\;
    }
    \Return $\hat{\kappa}$\;
 }
 \caption{Pseudocode for estimating $\kappa$}
 \label{algo:pseudokappa}
\end{algorithm}

We also derive a procedure for estimating $t$ given fixed $\kappa$ based on the idea of minimizing a distributional distance between the observed data distribution and that of theoretical tent pmfs generated based on a grid of candidate $t$ values and fixed $(\lambda,k)$. In simulation studies we have found that use of the Hellinger distance provides better estimates of $t$ in many settings compared to other choices, e.g. Chebyshev, Total Variation, or higher-power distances. The algorithm for selecting an estimate $\hat{t}$ for $t$ is provided in Algorithm \ref{algo:pseudot}.
\begin{algorithm}[t]
 \KwData{Observed data distribution $\hat{P}$, estimate $m$ of median $\lambda$, estimate of terminal node depth $\hat{d}$, estimate of scale prior mode hyperparameter $\hat{\kappa}$}
 \KwResult{An estimate $\hat{t}$ of $t$}
\tcc{obtain estimate $\hat{k}$ of $k$}
$\hat{k} \leftarrow \floor{\hat{\kappa}/2^{\hat{d}}}$\;
\tcc{generate grid of candidate $t$ values}
$t\_grid \leftarrow seq(0,0.49,by=increment)$\;
\tcc{generate theoretical tent pmfs $P_t(m,\floor{\exp(\hat{k})})$ over grid of candidate $t$ values}\;
$dists \leftarrow \{ P_t(m,\floor{\exp(\hat{k})}) : t \in t\_grid \}$\;
\tcc{compute Hellinger distances $H(\hat{P},P_t)$ and find minimizing $t$}
$\hat{t} \leftarrow \argmin_{t \in t\_grid} H(\hat{P},dists)$\;
\Return $\hat{t}$\;
 \caption{Pseudocode for estimating $t$ via distance minimization}
 \label{algo:pseudot}
\end{algorithm}

\begin{algorithm}[ht]
 \KwData{Observed data distribution $\hat{P}$, estimate $m$ of median $\lambda$, estimate of terminal node depth $\hat{d}$, maximum number of iterations $N$, stopping criteria $(\epsilon_{\kappa},\epsilon_t)$}
 \KwResult{Estimate $(\hat{\kappa},\hat{t})$ of $(\kappa,t)$}
Initialize $(t^{(0)},k^{(0)})$ (e.g. MLEs) and use $k^{(0)}$ to initialize $\kappa^{(0)}$\;
\For{i in 1:N}{
    Generate $t^{(i)}$ via Algorithm \ref{algo:pseudot}\;
    Generate $\kappa^{(i)}$ via Algorithm \ref{algo:pseudokappa}\;
    \eIf{ ($|t^{(i-1)} - t^{(i)}| < \epsilon_t$ and $|\kappa^{(i-1)} - \kappa^{(i)}| < \epsilon_{\kappa}$) or i==N}{
    Set $\hat{t} = t^{(i)}$\;
    Set $\hat{\kappa} = \kappa^{(i)}$\;
    Break\;
    }{$i++$\;}
}
\Return $(\hat{\kappa},\hat{t})$\;
 \caption{Pseudocode for estimating $\kappa$ and $t$}
 \label{algo:pseudokappat}
\end{algorithm}
Since the estimation procedure for $t$ depends on the observed data distribution $\hat{P}$, we recommend splitting the data into training and hold-out sets and using only the training data to derive $\hat{P}$.

Finally, Algorithms \ref{algo:pseudokappa} and \ref{algo:pseudot} suggest a joint estimation procedure for $(\kappa,t)$, specified in Algorithm \ref{algo:pseudokappat}. Note that the algorithm initializes $\kappa^{(0)}$ by setting $k$ equal to the first moment of the scale prior distribution, i.e. $k \approx \floor{\kappa / 2^d}$.

For calibration of the scale prior distribution scale hyperparameter $\beta_k$, we recommend restricting candidate values to the interval $[0,1]$ in cases where counts tend to be smaller, as the presence of the floor function and $\widetilde{\log}(\lambda)$ in the scale term tend to quickly reduce the scale prior scale value to zero at relatively shallow node depths for $\beta_k > 1$. In general, a cross-validation approach may be used to identify useful candidate hyperparameter values over a range of reasonable values.

\section{Zero-Inflated Extension}
\label{appx:zi-extension}

\subsection{ZI Model Setup}

We may also extend our proposed model to handle cases in which data exhibit excess zeros, relying on a typical mixture distribution formulation \citep[see e.g.][]{lambert_1992}. In this setting, we introduce latent variables $Z_i \simind \Bern (\rho_i)$ to form complete data $(Y,Z)$, such that $Y_i \giv Z_i=1 \sim \delta_{ \{ 0\} }$ and $Y_i \giv Z_i=0 \sim P_t(\lambda_i, \floor{e^{k_i}})$. As a prior distribution we specify $\rho_i \giv \mathcal{T} \simiid \text{Beta} (h_1,h_2)$, $(h_1,h_2) \in \R_{>0}^{2}$. The joint data distribution in this ZI setting is thus
\begin{align}
\label{eq:jointzidistn}
    p_t(\mathbf{Y},\mathbf{Z} \giv \bm{\lambda},\bm{k},\bm{\rho},\mathcal{T}) = \prod_{i=1}^n \rho_i^{Z_i}\left[ (1-\rho_i)p_t(y_i \giv \lambda_i,k_i) \right]^{1-Z_i},
\end{align}
and viewed as a likelihood in $(\bm\lambda,\bm k,\bm\rho)$ for fixed $\mathcal{T}$, each terminal node in this ZI model now contains three parameters $(\lambda_b,k_b,\rho_b)$, so that (\ref{eq:jointzidistn}) is rewritten as
\begin{align*}
    \Lik(\bm\lambda,\bm k,\bm\rho \giv \bfY,\bfZ,\cdot) &= \prod_{b=1}^B \Lik(\lambda_b,k_b,\rho_b \giv \bfY,\bfZ,\cdot) \\
    &= \prod_{b=1}^B \prod\limits_{\substack{i:y_i \in \eta_b \\ \wedge \ Z_i=1}} \rho_b \prod\limits_{\substack{i:y_i \in \eta_b \\ \wedge \ Z_i=0}} (1-\rho_b) p_t(Y_i \giv \lambda_b,k_b) \nonumber \\
    &= \prod_{b=1}^B \left[ \rho_b^{n_b^1} (1-\rho_b)^{n_b-n_b^1} \right] \prod\limits_{\substack{i:y_i \in \eta_b \\ \wedge \ Z_i=0}} p_t(Y_i \giv \lambda_b,k_b) \nonumber \\
    &= \prod_{b=1}^B \Lik(\rho_b \giv \bfZ_b,\cdot) \Lik(\lambda_b,k_b \giv \bfY_b,\bfZ_b,\cdot), \label{eq:jointzilikfactor}
\end{align*}
where we let $n_b = \#\{i:Z_i \in \eta_b \}$, $n_b^1 = \sum_{i:Z_i \in \eta_b} Z_i$, $\bm Y_b = \{Y_i : Y_i \in \eta_b \}$, $\bm Z_b = \{Z_i: Z_i \in \eta_b \}$, and suppress dependence of the above expressions on $\bf{X}$ for simplicity. The non-ZI likelihood is recovered by setting all $Z_i=0$ and dropping terms depending on each $\rho_b$.

The full hierarchical ZI model may now be specified as follows:
\begin{equation*}
\begin{aligned}
Y_i \giv (Z_i=0),t,\mathcal{T},\lambda_i,k_i &\simind P_t(\lambda_i,\floor*{\exp\{k_i \}}),  \\
Y_i \giv (Z_i=1),t,\mathcal{T},\cdot &\simind \delta_{\{ 0\}},\ i=1,\dots,n, \\
Z_i \giv \rho_i,\mathcal{T} &\simind \text{Bern}(\rho_i),\ i=1,\dots,n, \\
\lambda_b \giv d_1,d_2,\mathcal{T} &\simiid \text{DU}\{d_1,d_2 \},\ b=1,\dots,B \\
k_b \giv \lambda_b,\kappa,\beta_k,t_k,\mathcal{T} &\simind P_{t_k} \left( \floor*{\frac{\kappa }{2^{d(\eta_b)} } }, \floor*{ \frac{\widetilde{\log}( \lambda_b)}{(1+d(\eta_b))^{\beta_k}} } \right),\ b=1,\dots,B \\
\rho_b \giv h_1,h_2,\mathcal{T} &\simiid \text{Beta}(h_1,h_2),\ b=1,\dots,B, \\
\mathcal{T} \giv \alpha,\beta,\zeta_v &\sim \pi(\mathcal{T} \giv \alpha,\beta,\zeta_v).
\end{aligned}
\end{equation*}

We point out that we could alternatively model $\rho$ through a second, separate tree $\mathcal{T}_{\rho}$, with the framework for such a ``separate-tree" single-tree model requiring few technical adjustments relative to the ``shared-tree" approach presented in this subsection.

\subsection{Latent Variable and Parameter Updating in the ZI Model}
\label{ss:zrhoupdatezimodel}

We require update algorithms for the latent vector $\mathbf{Z}$ and the parameter vector $\bm\rho$ in the ZI model. The conditional distributions for the $Z_i$'s are straightforward in this case and follow immediately from the mixture distribution representation described in the previous subsection. Here we define $\M_b = (\lambda_b,k_b,\rho_b)$ and $\M_{b(-\rho)} = (\lambda_b,k_b)$ for compactness:

\label{eq:ztcondd1}
\[ p(Z_i=0 \giv Y_i=y_i,\M_b,\cdot) = \begin{cases*} 
      \frac{(1-\rho_b)p_t(Y_i=0|\M_{b(-\rho)},\cdot)}{\rho_b+(1-\rho_b)p_t(Y_i=0|\M_{b(-\rho)},\cdot)}, & \text{ if } $y_i=0$, \\
      1, & \text{ if } $y_i>0$, \\
        0, & \text{ otherwise }, \\
   \end{cases*}
\]
and 
\label{eq:ztcondd2}
\[ p(Z_i=1 \giv Y_i=y_i,\M_b,\cdot) = \begin{cases*} 
      \frac{\rho_b}{\rho_b+(1-\rho_b) p_t(Y_i=0|\M_{b(-\rho)},\cdot)}, & \text{ if } $y_i=0$, \\
        0, & \text{ otherwise }. \\
   \end{cases*}
\]
The conditional posterior distribution for each $\rho_b$ is written as
\begin{align*}
    \pi(\rho_b \giv \cdot) &\propto \mathcal{L}(\rho_b \giv Z_b,\cdot) \pi(\rho_b \giv \mathcal{T}) \\
    &\propto \rho_b^{n_b^1} (1-\rho_b)^{n_b-n_b^1} \times \rho_b^{h_1-1}(1-\rho_b)^{h_2-1} \\
    &= \rho_b^{h_1 -1 + n_b^1} (1-\rho_b)^{h_2-1 + (n_{b}-n_b^1)},
\end{align*}
which we recognize as the kernel of a $\text{Beta}(h_1 + n_b^1,h_2+n_b-n_b^1)$ distribution, so that each $\rho_b$ may be updated using a Gibbs step.

\subsection{The $\rho$-Marginalized Likelihood}

Importantly, since the conditional posterior for $\rho_b$ is an exponential family distribution, we can simply integrate out each $\rho_b$ from the joint function $\mathcal{L}(\eta_b \giv \bfY_b,Z_b,\T,\M_b)\pi(\rho_b \giv \T)$ and work instead with the $\rho$-marginalized likelihood contribution from each terminal node $\eta_b$, represented as $\mathcal{L}(\eta_b \giv \bfY_b,Z_b,\T,\M_{b(-\rho)})$, when evaluating tree birth and death proposals in the ZI model:
\begin{align*}
    \mathcal{L}(\eta_b &\giv \bfY_b,Z_b,\T,\M_{b(-\rho)}) \\
    &= \int_{\rho_b} \mathcal{L}(\eta_b\giv\bfY_b,Z_b,\T,\M_b)\pi(\rho_b\giv\T) \ d \rho_b \\
    &= \int_{\rho_b} \rho_b^{n_b^1} \prod_{\substack{i:Y_i \in \eta_b, \\ Z_i=0}} (1-\rho_b)p(Y_i|\M_{b(-\rho)}) \times \frac{\Gamma(h_1+h_2)}{\Gamma(h_1)\Gamma(h_2)} \rho_b^{h_1-1}(1-\rho_b)^{h_2-1} \ d \rho_b\\
    &= \prod_{\substack{i:Y_i \in \eta_b, \\ Z_i=0}} p_t(Y_i|\M_{b(-\rho)})\times \frac{\Gamma(h_1+h_2)}{\Gamma(h_1)\Gamma(h_2)} \int_{\rho_b} \rho_b^{h_1 - 1 + n_b^1}(1-\rho_b)^{h_2-1+n_b-n_b^1} \ d \rho_b\\
    &= \frac{\Gamma(h_1+h_2)\Gamma(h_1 + n_b^1)\Gamma(h_2+n_b-n_b^1)}{\Gamma(h_1)\Gamma(h_2) \Gamma(h_1 + h_2 + n_b)} \times \prod_{\substack{i:Y_i \in \eta_b, \\ Z_i=0}} p_t(Y_i \giv \M_{b(-\rho)}).
\end{align*}

\subsection{Birth and Death Move Proposals in the ZI Single-Tree Count Model}

Acceptance probability calculations for tree birth and death move proposals in the ZI single-tree count model are handled in a similar fashion to the non-ZI case, with the difference being that the likelihood ratio (LR) component of the MH ratio is now based on the $\rho$-marginalized likelihood, i.e.

\begin{equation*}
    \text{LR} = \frac{ \mathcal{L}(\eta_{b(l)}\giv\bfY_{b(l)},\bfZ_{b(l)},\bfM_{b(l)(-\rho)})\mathcal{L}(\eta_{b(r)}\giv\bfY_{b(r)},\bfZ_{b(r)},\bfM_{b(r)(-\rho)}) }{ \mathcal{L}(\eta_b\giv\bfY_b,\bfZ_b,\bfM_{b(-\rho)}) }.
\end{equation*}

\subsection{ZI Runtime Comparison Example}

In this subsection we fit the ZI model using the TC and MH samplers to data generated from a true data-generating process contaminated by an excess zero component and attempt to recover the correct tree structure and terminal node parameters. In this ZI example we considered the following two-covariate true data-generating model:
\begin{align*}
    Y_i &\giv Z_i=0,\bfX_{i} \stackrel{ind.}{\sim} P_t(g_1(\bfX_i),\floor{\exp(g_2(\bfX_i))}), \\
    Y_i &\giv Z_i=1,\bfX_{i} \stackrel{ind.}{\sim} \delta_{\{ 0\}}, \\
    Z_i &\giv \bfX_{i} \stackrel{ind.}{\sim} \text{Bern}(g_3(\bfX_i)), \label{eq:runtimeextruezimodel}
\end{align*}
for $i=1,\dots,n$, where letting $g(\bfX_i) = (g_1(\bfX_i),g_2(\bfX_i),g_3(\bfX_i))$, we define
\begin{equation*}
 g(\bfX_i) = \begin{cases} 
      (2,1,0.3) , &  x_{i1} \leq 5, x_{i2} \leq 5 \\
      (3,1,0) , &  x_{i1} \leq 5, x_{i2} > 5 \\
      (1,0,0) , &  x_{i1} > 5, x_{i2} \leq 5 \\
      (7,2,0.2) , &  x_{i1} > 5, x_{i2} > 5,
   \end{cases}
\end{equation*}

with $X_{ij} \sim \text{Unif}(0,10)$ for $i = 1,...,n$ and $j=1,2$. For simplicity we once more took $t=0$ in this true model. We proceeded to generate $n=10,\ 100,$ and $1000$ observations according to this true model for use in our fitting procedures. Table \ref{table:runtimezidatasummary} displays select summary information from the generated counts at each sample size setting.

\begin{table}[!htbp]
\centering
\begin{tabular}{@{} cccc @{}}
  \toprule
    \textbf{n} & \textbf{\% Zeroes} & \textbf{Mean} & \textbf{5-Number Summary} \\
  \midrule
  10 & 30\% & 3.9 & (0,1,2,7,12) \\
  100 & 21\% & 2.6 & (0,1,2,3.3,14) \\
  1000 & 21.6\% & 2.7 & (0,1,2,4,14) \\
  \bottomrule
\end{tabular}
\caption{Select summary statistics for the generated counts at each sample size setting in the zero-inflated simulation study. All reported values are rounded to one significant digit.}
\label{table:runtimezidatasummary}
\end{table}

\subsubsection{Setup}

For this experiment, the dimension-changing proposals for the ZI model under the TC sampler follows an analogous construction to that of the birth/death moves described in the main manuscript as specified in the previous subsections. The full updating algorithm for our proposed single-tree ZI count model using the taxicab sampler is described in Algorithm \ref{algo:fullzimodelupdate}.

\begin{center}
\begin{algorithm}[H]
 \KwData{Realized observations $(Y_1,\mathbf{X}_1),\dots,(Y_n,\mathbf{X}_n)$ }
 \KwResult{Approximate posterior samples drawn from $\pi(\bm{\lambda},\bm{k},\bm{\rho},\mathcal{T},\bfZ, \mathbf{U}, \mathbf{R} \giv (Y_1,\mathbf{X}_1),\dots,(Y_n,\mathbf{X}_n)$)}
 \For{$N_{mcmc}$ iterations}{
	Update $\bfZ \giv \cdot$ via a Gibbs step \\
	Propose $(\mathcal{T}',\tilde{\bm{\lambda}},\tilde{\bm{k}},\tilde{\mathbf{U}},\tilde{\mathbf{R}}) \giv \cdot \sim q(\mathcal{T}',\tilde{\bm{\lambda}},\tilde{\bm{k}},\tilde{\mathbf{U}},\tilde{\mathbf{R}} \giv \mathcal{T},\bm{\lambda},\bm{k},\mathbf{U},\mathbf{R},\bfZ)$ and accept/reject via a dimension-changing MH step using the $\rho$-marginalized likelihood \\
    Draw $(\bm{\lambda}',\mathbf{U}') \giv \tilde{\bm{\lambda}},\tilde{\mathbf{U}},\cdot$ and $(\bm{k}',\mathbf{R}') \giv \tilde{\bm{k}},\tilde{\mathbf{R}},\cdot$ via TC sampler steps \\
	Update $\bm{\rho} \giv \cdot$ via a Gibbs step
}
 \caption{Posterior sampling algorithm for the proposed single-tree ZI count model with taxicab sampler}
 \label{algo:fullzimodelupdate}
\end{algorithm}
\end{center}

Again using the CGM98 ``restart" strategy, we ran 20 individual chains for 5000 iterations each and restarting each new chain from a single-node tree. 100 burn-in iterations were used for each run and discarded prior to analysis. We utilized $\zeta_v=50$ cuts to discretize each covariate dimension. Hyperparameter settings for this set of comparison simulations are detailed in Table \ref{table:runtimehpzi}. 
\begin{table}[!htbp]
\centering
\begin{tabular}{@{} ccc @{}}
  \toprule
    \textbf{Method} & \textbf{Parameters} & \textbf{Values Considered} \\
  \midrule
    \multirow{4}{*}{Na{\"i}ve MH} & $k$ prior: $(\kappa, \beta_k, t_k)$ combinations & (2,0,0.025)  \\
     & Tree depth prior: $(\alpha, \beta)$ combinations & (0.95,2) \\
 & MH proposal radii: $(\lambda,k,c)$ combinations & (4,2,25), (6,2,25) \\
 & Beta prior: $(h_1,h_2)$ combinations & (1,1) \\
 & Tent pmf tail mass parameter: $t$ & 0.025 \\
 \midrule
    \multirow{4}{*}{Taxicab} & $k$ prior: $(\kappa, \beta_k,t_k)$ combinations & (2,0,0.025) \\
 & Tree depth prior: $(\alpha, \beta)$ combinations & (0.95,2) \\
 & $\mathcal{N}(\cdot)$ radii: $(m_{\lambda},m_k)$ combinations & (2,1), (3,1), (4,2), (5,2) \\
 & MH proposal radius: $c$ & 25 \\
  & Beta prior: $(h_1,h_2)$ combinations & (1,1) \\
 & Tent pmf tail mass parameter: $t$ & 0.025 \\
  \bottomrule
\end{tabular}
\caption{Hyperparameter settings for runtime comparison.}
\label{table:runtimehpzi}
\end{table}
In the scale prior distribution we chose the hyperparameter $\beta_k=0$ to allow for greater spread in the prior distribution over $k$ values near the prior mode. We used a noninformative $\text{Beta}(1,1)$ prior distribution for the success probabilities $\bm\rho$. Here we took $t=0.025$ and $t_k = 0.025$ for all runs. Assessment of fit was based on mean absolute error (MAE) and $L_2$ norm, both averaged over the 20 runs at each combination of hyperparameter settings. 

We took 1000 posterior samples to compute both the $L_2$ norm and MAE quantities, along with their standard deviation (SD) and standard error (SE) respectively. Total runtime was also recorded at each combination of model settings, measuring the length of time elapsed to execute the model-fitting algorithm for all 20 runs. The results of this comparison are presented in Tables \ref{table:ziresults1} and \ref{table:ziresults2}, corresponding to the respective outcomes for the na{\"i}ve MH and TC approaches.
\begin{table}[ht]
\centering
\begin{tabular}{@{} cccc @{}}
  \toprule
    \textbf{Method} & \textbf{n} & \textbf{$(\lambda,k,c)$ radii} & \textbf{Runtime(sec)} \\
  \midrule
    \multirow{6}{*}{Na{\"i}ve MH} & 10 & (4,2,25) & 54.19 \\
    & 10 & (6,2,25) & 53.77 \\ 
    & 100 & (4,2,25) & 261.97 \\
    & 100 & (6,2,25) & 258.45 \\
    & 1000 & (4,2,25) & 1942.54 \\
    & 1000 & (6,2,25) & 1951.63 \\
   \midrule
     \textbf{Method} & \textbf{n} & \textbf{$(m_{\lambda},m_k,c)$ radii} & \textbf{Runtime(sec)} \\
   \midrule
    \multirow{12}{*}{Taxicab} & 10 & (2,1,25) & 11.80 \\
    & 10 & (3,1,25) & 13.81 \\
    & 10 & (4,2,25) & 21.85 \\ 
    & 10 & (5,2,25) & 23.31 \\ 
    & 100 & (2,1,25) & 64.53 \\
    & 100 & (3,1,25) & 74.05 \\
    & 100 & (4,2,25) & 102.11 \\
    & 100 & (5,2,25) & 107.82 \\
    & 1000 & (2,1,25) & 517.47 \\
    & 1000 & (3,1,25) & 603.76 \\
    & 1000 & (4,2,25) & 814.69 \\
    & 1000 & (5,2,25) & 872.39 \\
  \bottomrule
\end{tabular}
\caption{Comparison of runtime results for ZI models fit with na{\"i}ve MH and TC approaches. All reported values are rounded to the nearest hundredths place.}
\label{table:ziresults1}
\end{table}
\begin{table}[ht]
\centering
\begin{tabular}{@{} ccccc @{}}
  \toprule
    \textbf{Method} & \textbf{n} & \textbf{$(\lambda,k,c)$ radii} & \textbf{MAE(SE)} & \textbf{$L_2$ norm(SD)} \\
  \midrule
    \multirow{6}{*}{Na{\"i}ve MH} & 10 & (4,2,25) & 2.53(0.01) & 51.41(19.34) \\
    & 10 & (6,2,25) & 2.55(0.01) & 51.47(17.71) \\ 
    & 100 & (4,2,25) & 1.39(0.00) & 31.81(4.62) \\
    & 100 & (6,2,25) & 1.39(0.00) & 31.28(3.93) \\
    & 1000 & (4,2,25) & 1.58(0.00) & 28.38(1.37) \\
    & 1000 & (6,2,25) & 1.58(0.01) & 28.55(2.4) \\
  \midrule
   \textbf{Method} & \textbf{n} & \textbf{$(m_{\lambda},m_k,c)$ radii} & \textbf{MAE(SE)} & \textbf{$L_2$ norm(SD)} \\
  \midrule 
    \multirow{12}{*}{Taxicab} & 
    10 & (2,1,25) & 2.51(0.01) & 52.43(15.24) \\
    & 10 & (3,1,25) & 2.51(0.00) & 53.05(17.51) \\
    & 10 & (4,2,25) & 2.45(0.01) & 52.78(19.61) \\ 
    & 10 & (5,2,25) & 2.46(0.01) & 52.04(20.02) \\ 
    & 100 & (2,1,25) & 1.37(0.01) & 30.58(3.36) \\
    & 100 & (3,1,25) & 1.37(0.01) & 30.26(3.41) \\
    & 100 & (4,2,25) & 1.36(0.00) & 30.33(4.00) \\
    & 100 & (5,2,25) & 1.36(0.00) & 29.50(3.04) \\
    & 1000 & (2,1,25) & 1.58(0.00) & 28.26(0.63) \\
    & 1000 & (3,1,25) & 1.58(0.00) & 28.31(0.77) \\
    & 1000 & (4,2,25) & 1.57(0.00) & 28.1(0.81) \\
    & 1000 & (5,2,25) & 1.57(0.00) & 28.06(0.84) \\
  \bottomrule
\end{tabular}
\caption{Comparison of MAE and $L_2$ norm results for ZI models fit with na{\"i}ve MH and TC approaches. All reported values are rounded to the nearest hundredths place.}
\label{table:ziresults2}
\end{table}

\subsubsection{Results}

Performance between the two samplers with respect to both $L_2$ norm and in-sample MAE is once again comparable across sample sizes, with improved recovery of the true tree structure and parameters as $n$ increases. In this case, the TC sampler was anywhere between 2 and 4 times faster than the MH sampler for ``similar" $\mathcal{N}(\cdot)$ and proposal radii settings. Relative to the non-ZI example, here we see smaller runtime improvements under the TC sampler compared to the MH sampler since the prior distribution support for $\lambda$ is far smaller in this case than in the non-ZI simulation study, thus requiring fewer total sweeps to numerically marginalize over $\lambda$ and $k$ when computing the marginal log likelihood at each iteration of a chain in tree birth and death proposals. In this case, for fixed $n$ the differences in MAE and $L_2$ norm at different $\mathcal{N}(\cdot)$ radii settings are effectively negligible under the TC sampler, suggesting that smaller choices of radii are sufficient for good exploration of the conditional posterior distributions for $\lambda$ and $k$.
\end{appendices}

\end{document}